\documentclass[onecolumn,showpacs,showkeys,preprintnumbers,aps]{revtex4}
\usepackage{graphicx}% Include figure files
\usepackage{dcolumn}% Align table columns on decimal point
\usepackage{bm}% bold math
\usepackage{epsfig}
\usepackage{longtable}
\begin{document}

\title{Spectroscopy of $^{230}$Th in  the (p,t) reaction}

\author{A.I.~Levon$^{1,*}$, G.~Graw$^2$, Y.~Eisermann$^2$, R.~Hertenberger$^2$,
 J.~Jolie$^3$,  N.Yu.~Shirikova$^4$, A.E.~Stuchbery$^5$, A.V.~Sushkov$^4$, P.~G.~Thirolf$^2$,  H.-F.~Wirth$^2$, N.V.~Zamfir$^6$}

\affiliation{$^1$ Institute for Nuclear Research, Academy of
Science, Kiev, Ukraine}

\affiliation{$^2$ Fakult\"at f\"ur Physik,
Ludwig-Maximilians-Universit\"at M\"unchen, Garching, Germany}

\affiliation{$^3$ Institut f\"ur Kernphysik, Universit\"at zu
K\"oln,  K\"oln, Germany}

\affiliation{$^{4}$  Joint Institute for Nuclear Research, Dubna,
Russia}

\affiliation{$^{5}$ Department of Nuclear Physics, Australian
National University, Canberra, Australia}

\affiliation{$^6$ H.~Hulubei National Institute of Physics and
Nuclear Engineering, Bucharest, Romania}

\email[ Electronic address: ] {levon@kinr.kiev.ua}

\date{\today}

\begin{abstract}
The excitation spectra in the deformed nucleus $^{230}$Th were
studied by means of the (p,t) reaction, using the Q3D spectrograph
facility at the Munich Tandem accelerator. The angular
distributions of tritons
 are measured for about 200 excitations seen in the triton spectra up
 to 3.3 MeV.  Firm $0^+$ assignments are made for 16 excited states by
 comparison of experimental angular distributions with the calculated
 ones using the CHUCK code. Additional assignments are possible:
 relatively firm for 4 states and tentative also for 4 states. Assignments
 up to spin $6^+$  are made for other states. Sequences of the states
 are selected which can be treated as rotational bands and as multiplets
of excitations. Experimental
 data are compared with  interacting boson model (IBM)  and
 quasiparticle-phonon model (QPM) calculations.
\end{abstract}


\maketitle
%%%%%%%%%%%%%%%%%%%%%%%%%%%%%%%%%%%%%%%%%%%%%%%%%%%%%%

\section{Introduction}
%\noindent
A full microscopic description of low-lying excitations in
deformed nuclei has eluded theoretical studies to date. Along with
the interplay of collective and single-particle excitations, which
takes place in deformed rare earth nuclei, additional problems
arise in the actinide region because of the reflection asymmetry
\cite{But96}. Evidently the nature of the first excited 0$^+$
states in the actinide nuclei is different from that in the rare
earth region where they are due to the quadrupole vibration
\cite{Mah72}. Octupole degrees of freedom have to be important in
the actinides. One has  then to expect a complicated picture at
higher excitations: residual interactions could mix the one-phonon
and multiphonon vibrations of quadrupole and octupole character
with each other and with quasiparticle excitations. Detailed
experimental information on the properties of such excitations is
needed for comparison with theory. On the experimental side, the
(p,t) reaction is very useful. On the theoretical side, a
microscopic approach such as the quasiparticle-phonon model (QPM)
is necessary, in order to account for the number of states
detected and to make detailed predictions on their properties .

Our previous paper \cite{Wir04}  concentrated on the
 excited $0^+$ states  in the actinide nuclei  $^{228}$Th, $^{230}$Th
 and $^{232}$U studied in the (p,t) reaction. This interest
 in the $0^+$ excitations in the deformed
 actinide nuclei was stimulated by  the observation of  multiple
 $L=0$ transfers in the (p,t) reaction leading to the excited states in
 the odd nucleus $^{229}$Pa \cite{Lev94} as well as to the excited states in
 the medium heavy even nuclei  $^{146}$Nd \cite{Pon96}, $^{146}$Sm
 \cite{Oro97}, $^{132,134}$Ba \cite{Cat96}.
After comprehensive information on 0$^+$ excitations was obtained
in the actinide nuclei \cite{Wir04}, systematic studies of the
same type were made on many nuclei between $^{152}$Gd and
$^{192}$Hg \cite{May05} and revealed a large number of excited
0$^+$ states, whose nature is not yet understood.
 The (p,t) reaction, however,
 gives much more extensive information on  specific excitations in these
 nuclei, which was not analyzed previously \cite{Wir04}.
 An attempt to obtain such information was made for $^{168}$Er after using
  a high-resolution experimental study with the (p,t) reaction
\cite{Buc06}. In this paper we present  the results of
 a careful and detailed analysis of the experimental data from
 the high-precision, high-resolution study of
the $^{232}$Th(p,t)$^{230}$Th reaction carried out  to get deeper
insight into excitations in $^{230}$Th including the nature of the
$0^+$ excitations.
 %%%%%%%%%%%%%%%%%%%%%%%%%%%%%%%%%%%%%%%%%%%%
\begin{figure*}
\begin{center}
\epsfig{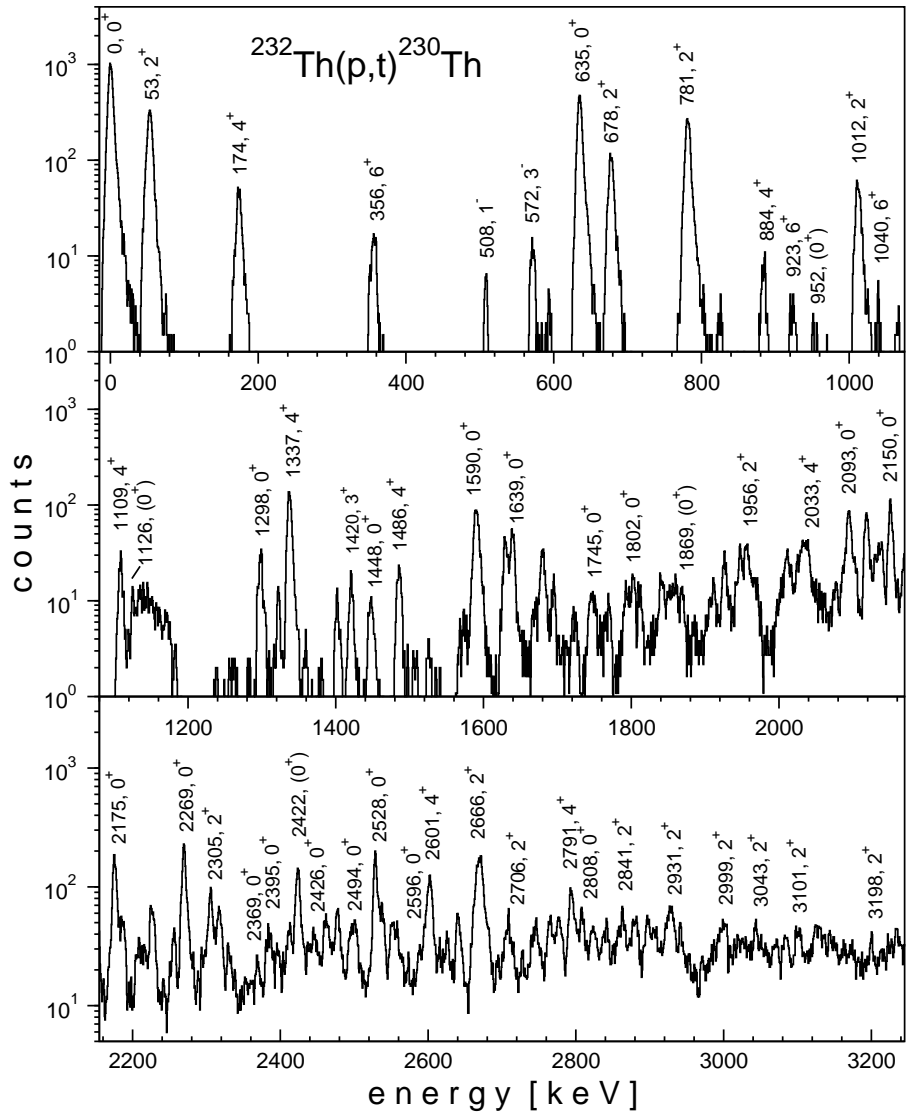}
\caption{\label{fig:specTh230_75deg} Triton energy spectrum from
the $^{232}$Th(p,t)$^{230}$Th reaction (E$_p$=25 MeV) in
 logarithmic scale for a detection angle of 7.5$^\circ$. Some strong lines
 are labeled with the   corresponding level energies in keV and by
the spins assigned from the DWBA fit.}
\end{center}
\end{figure*}
%%%%%%%%%%%%%%%%%%%%%%%%%%%%%%%%%%%%%%%%%%%%%%%%%

Experimentally accessible thorium nuclei span a wide and
interesting
 range of isotopes: from the semi-magic $^{216}$Th  to
quadrupole-deformed $^{234}$Th. An especially interesting region
is
 that around A=228, where the even isotopes $^{226}$Th, $^{228}$Th and
 $^{230}$Th are considered to be octupole-deformed, octupole soft and
 vibrational-like, respectively \cite{Egi89}. Two of them are accessible
 for study by the (p,t) reaction. Besides $0^+$ excitations, whose
 number  is increased in comparison with the preliminary analysis
 in publication \cite{Wir04}, the spins for many other states are also
 assigned. The results concerning 0$^+$, 2$^+$ and 4$^+$ excitations are
 mainly discussed. The 6$^+$ and negative parity states are excited
 weakly in the (p,t) reaction; therefore information on these states
 is only fragmentary. The results of a similar analysis for $^{228}$Th
 in comparison
 with $^{229}$Pa as well as for $^{232}$U will be presented in
 forthcoming papers \cite{Levon}.

 In Sec.~\ref{det_exp} of this paper the details of the
 experimental techniques and the experimental results are given.

 After description of the DWBA analysis a preliminary treatment of
 the nature of some excitations is presented. In Sec.~\ref{Disc}
 the interpretation of 0$^+$ excitations is
 discussed. The experimental data obtained in this study are
 compared with the results of calculations in microscopic approach
 of the quasiparticle-phonon model (QPM). The validity  of the phenomenological
 approach of the interacting boson model (IBM) is also tested to account
 for at least the main features  of the positive parity excitations in
 $^{230}$Th, in addition to $0^+$ states tested in \cite{Wir04}.

\section{Experiment, analysis and experimental data}

\subsection{\label{det_exp} Details of the experiment}
%%%%%%%%%%%%%%%%%%%%%%%%%%%%%%%%%%%%%%%%%%%%%%%%%%%%%%%%%%%%%%%%%%%%
\begin{figure*}
\begin{center}
\epsfig{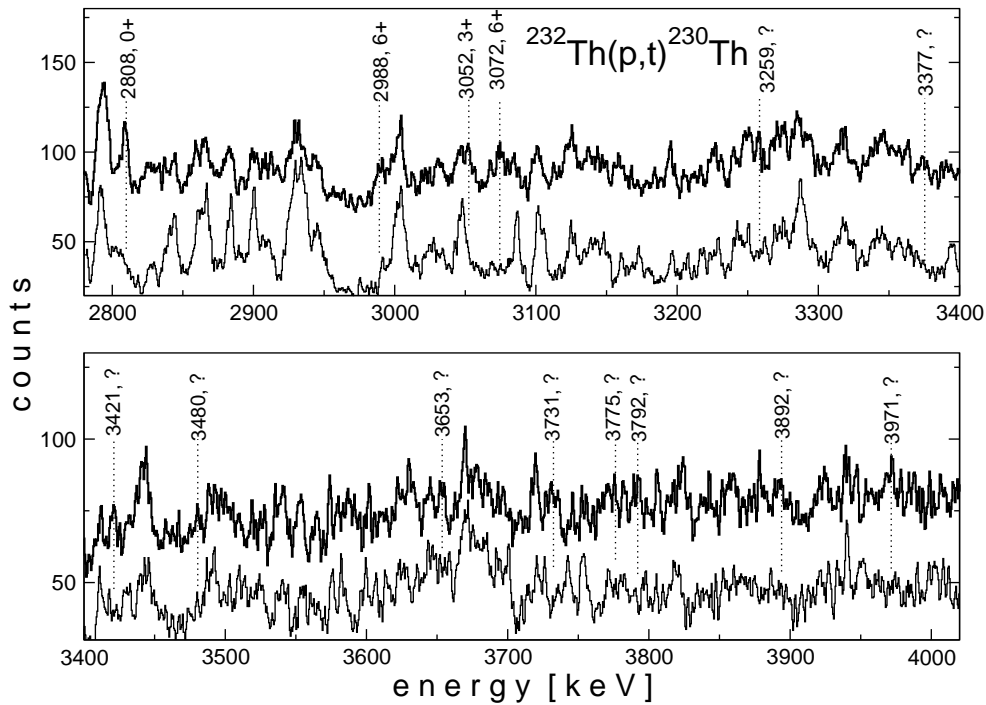}
\caption{\label{fig:spec_high} Comparison of the spectra for the
  $^{232}$Th(p,t)$^{230}$Th reaction
 (E$_p$=25 MeV) for detection angles of 12.5$^\circ$ (thin line) and 26$^\circ$
 (thick line) for the high energy range. The lines corresponding to possible $0^+$
 excitations are labeled by their energies.}
\end{center}
\end{figure*}
%%%%%%%%%%%%%%%%%%%%%%%%%%%%%%%%%%%%%%%%%%%%%%%%%%%%%%%%%%%%%

A target of  100 $\mu$g/cm$^2$ $^{232}$Th evaporated on a 22
 $\mu$g/cm$^2$ thick carbon backing was bombarded with 25 MeV protons
of an intensity of 1-2 $\mu$A from the Tandem accelerator of the
Maier-Leibnitz-Labor of the Ludwig-Maximilians-Universit\"at and
Technische Universit\"at M\"unchen. The isotopic purity of the
target was about 99\,\%. The tritons were analyzed with the Q3D
magnetic spectrograph  and then
 detected in a focal plane detector. The focal plane detector is a multiwire
 proportional chamber with readout of a cathode foil structure for
 position determination and dE/E particle identification
 \cite{Zan91,Wir01}. The acceptance of the spectrograph was 11 msr,
except for the most forward angle of 5$^\circ$ with an acceptance
of 6 msr.
 The resulting triton spectra have a resolution of
 4--7 keV (FWHM) and are background free for all
 angles but 5$^\circ$ for which background from light
 contaminations
 in the region of  1100-1150 keV complicated the analysis  for some levels.
 The angular distributions of the cross sections were obtained from the
 triton spectra at ten laboratory angles from 5$^\circ$ to 45$^\circ$.

 A triton energy
spectrum measured at a detection angle of 7.5$^\circ$ is shown in
 Fig.~\ref{fig:specTh230_75deg}, which demonstrates the domination of $0^+$
 excitations having large cross sections at this angle. The analysis of
 the triton spectra was performed with the program GASPAN \cite{Rie91}.
Measurements were carried out with two magnetic settings: one for
 excitations up to 1.6 MeV and another for the energy region
 from 1.5 MeV to 3.3 MeV.   For the calibration of the energy scale
the triton spectra from the reaction $^{184}$W(p,t)$^{182}$W,
 $^{186}$W(p,t)$^{184}$W  and $^{234}$U(p,t)$^{232}$U were
 measured at the same magnetic settings. Some known levels
in $^{230}$Th were included in the calibration, leading to small
 corrections from 0 to 0.5 keV for excitations between 1.5  MeV and
3.3 MeV. In the course of our measurements we found that the Q
values for the (p,t) reactions on the $^{232}$Th, $^{230}$Th and
$^{234}$U targets or on the $^{186}$W and $^{184}$W targets are in
disagreement with the ones calculated from the Atomic Mass
Evaluation \cite{Aud03} depending which Q value is used as the
reference value. They are given in Table~\ref{tab:Q}. Notation
"Ref." (the reference value) in this table means that the Q value
for this nucleus as determined from the data in the AME was taken
as a starting point in the calculations for other nuclei. If the Q
value for the $^{232}$Th(p,t) reaction as derived from
\cite{Aud03} is taken as the reference value, then the Q values
for $^{230}$Th(p,t) and $^{234}$U(p,t) reactions also agree with
those calculated from the AME table \cite{Aud03}. At the same time
 the Q values for the $^{186}$W(p,t) and $^{184}$W(p,t) reactions
differ considerably from the calculated one and vice versa, if the
Q value for the $^{184}$W(p,t) reactions is taken as the reference
value. It is not clear from these data what is the reason for this
discrepancy.

About 200 levels have been identified in the spectra for all ten
angles and are listed in Table~\ref{tab:expEI}. The energies and
spins of the levels as derived from this study are compared to
known energies and spins from the published data
\cite{Ako93,Ack93,Ack94}. They are given in the first four
columns. The ratios of cross sections at angles 5$^\circ$ and
26$^{\circ}$ to that at angle 16$^{\circ}$, given in the next two
columns, help to highlight the $0^+$ excitations (large values).
The column $\sigma_{\mbox{integ.}}$ gives the cross section
integrated in the region from 5$^{\circ}$ to 45$^{\circ}$ and the
column {$\sigma_{\mbox{exp.}}/\sigma_{\mbox{calc.}}$} gives the
ratio of the integrated cross sections, from experimental values
and calculations in the DWBA approximation (see Sec.
\ref{sec:DWBA}). The last column gives the notations of the
schemes used in the DWBA calculations: sw.jj means one-step direct
transfer of the $(j)^2$ neutrons in the (p,t) reaction; notations
of the multi-way transfers used in the DWBA calculations are
displayed in Fig.~\ref{fig:schemes} (see Sec. \ref{sec:DWBA}).

\newcolumntype{d}{D{.}{.}{2}}
\begin{table}
\caption {\label{tab:Q} Q values for the (p,t) reactions as
determined from a comparison of the experimental triton spectra
and as derived from the AME 2003 compilation \cite {Aud03}.}
\footnotesize
 \begin{ruledtabular}
 \begin{tabular}{cccc}
 \smallskip\\
 Reaction & Ref.: $^{232}$Th(p,t) & Ref.: $^{184}$W(p,t) & AME 2003
 \cite{Aud03}\\
          & [keV] & [keV] & [keV]
   \smallskip\\
 \hline
\smallskip\\
$^{232}$Th(p,t) & -3076.5(27) & -3064.2(20) & -3076.5(27)\\
$^{230}$Th(p,t) & -3568.5(30) & -3556.6(20) & -3569.0(28)\\
$^{234}$U(p,t) & -4125.2(30) & -4113.0(20) & -4124.2(29)\\
$^{186}$W(p,t)  & -4478.1(30) & -4460.5(20) & -4463.0(19)\\
$^{184}$W(p,t)  & -5133.0(30) & -5120.6(12) & -5120.6(12)\\
   \end{tabular}
\end{ruledtabular}
\end{table}
%%%%%%%%%%%%%%%%%%%%%%%%%%%%%%%%%%%%%%%%%%%%%%%%%%%%%%%%%%%%%%%%%%%
\normalsize

In order to get an indication of the possible $0^+$ excitations at
higher energies, the triton spectra were measured for the angles
of 12.5$^\circ$ and
 26.0$^\circ$ and in the energy range from 2.8 MeV to 4.5 MeV.  A precise
 calibration of this region of excitations is not possible with
the spectra measured for  the tungsten and uranium targets or any
other targets. Therefore the calibration makes use of that
employed for the energy region 1550 - 3300 keV with a linear shift
of the energy and channel number determined from the position of
the known levels observed in both energy regions. The measured
spectra are compared in Fig.~\ref{fig:spec_high}. The ratio of the
line intensities
 at 2808 keV in two spectra can serve as a reference for the 0$^+$
state assignments. Peaks that may correspond to the excitation of
0$^+$ states are marked by dashed lines.
 Unfortunately, close values of the excitation ratios are
valid also for 6$^+$ and 3$^+$ excitations as shown by the states
at 3052 and 3072 keV. Nevertheless, at least 10 candidates for
0$^+$ excitations are revealed in the region from 3 to 4 MeV.
Measurements at the angle of 5$^\circ$ could confirm or reject
these assumptions.
%%%%%%%%%%%%%%%%%%%%%%%%%%%%%%%%%%%%%%%%%%%%%%%%%%%%%%%%%%%%%%%%%%
\newcolumntype{d}{D{.}{.}{3}}
\renewcommand{\thefootnote}{\fnsymbol{footnote}}
\begin{longtable*}{ll ll rcc rr r c c}
\caption{\label{tab:expEI} \normalsize Energies of  levels in
$^{230}$Th, the level spin assignments from the CHUCK analysis,
the (p,t) cross sections integrated over the measured values and
the reference to the schemes used in the DWBA calculations
(see text for more detailed explanations).}\\
\hline\hline
\smallskip\\
\multicolumn{4}{c}{Level energy
[keV]}&\multicolumn{2}{c}{$I^\pi$}&&\multicolumn{2}{c} {Cross
section ratios}& \enspace {$\sigma_{\mbox{integ.}}$}&{Ratio}&Way
of
\smallskip\\
\multicolumn{2}{l}{This work}&\multicolumn{2}{l}{[16-18]}&[16-18]
&This work&&{(5$^o$/16$^o$)} &{(26$^o$/16$^o$)}&[{$\mu$b}]&
{$\sigma_{\mbox{expt.}}/\sigma_{\mbox{calc.}}$}&fitting\\
\smallskip\\
\hline
%\smallskip
\endfirsthead
\caption{Continuation}\label{tab:expEI}\\
\hline\hline
\smallskip\\
\multicolumn{4}{c}{Level energy
 [keV]}&\multicolumn{2}{c}{$I^\pi$}&&\multicolumn{2}{c}
{Cross section ratios}& \enspace
{$\sigma_{\mbox{integ.}}$}&{Ratio}&Way of
\smallskip\\
%\cline{1-2}   \cline{3-4} \cline{6-7}
%\smallskip\\
\multicolumn{2}{l}{This work}&\multicolumn{2}{l}{[16-18]}&[16-18]&
This work&&{(5$^o$/16$^o$)} &{(26$^o$/16$^o$)}&[{$\mu$b}]&
{$\sigma_{\mbox{expt.}}/\sigma_{\mbox{calc.}}$}&fitting\\
\smallskip\\
\hline
%\smallskip
\endhead
\hline
\endfoot
\endlastfoot
~~~0.1 \it2&&    ~~~0.0 &&$0^+$&$0^+$&&             8.02 &   5.51 &  195.68 &10.7  &sw.gg\\
~~53.2 \it2 &&   ~~53.20 \it2&&$2^+$&$2^+$&&        1.57 &   0.27 &  52.53  &11.0 &m1a \\
~174.0 \it2&&    ~174.10 \it3&&$4^+$&$4^+$&&        0.87 &   0.50 &   9.94  &2.6  & m1a\\
 ~356.3 \it2 &&   ~356.6 \it5 &&$6^+$&$6^+$&&        0.40&   0.75 &   7.39  &2.8  & m2d\\
 ~508.0 \it3 &&   ~508.15 \it3&&$1^-$&$1^-$&&        0.63&   0.00 &   0.87  &0.6 & m2a\\
 ~571.7 \it2 &&   ~571.73 \it3 &&$3^-$&$3^-$&&       0.33 &  0.44 &   4.04  &0.8   & m3a\\
 ~593.8 \it3 &&   ~594.1 \it5   &&$8^+$&$8^+$&&      1.05 &   0.00 &    0.37 &0.4&m2c\\
 ~635.1 \it2 &&   ~634.88 \it5 &&$0^+$&$0^+$&&       26.91 &    8.92 &   47.79  & 250 & sw.ii\\
 ~677.6 \it2 &&   ~677.54 \it5 &&$2^+$&$2^+$&&        0.85 &    0.55 &   25.13  & 3.5 & m1a\\
 ~686.0 \it10  && ~686.7       &&$5^-$&     &&             &         &   $<0.2$ &     &\\
 ~775.2 \it4 &&   ~775.5 \it3  &&$4^+$&$4^+$&&        0.30 &   0.53 &    7.15   &1.3 & m1a \\
 ~781.4 \it2 &&   ~781.35 \it3 &&$2^+$&$2^+$&&        0.62 &    0.46 &   69.07 & 6.2 & sw.gg\\
 ~825.6 \it3 &&   ~825.66 \it5 &&$3^+$&$3^+$&&        0.14  &   0.79  &   1.10 & 2.5     & m2a\\
 ~852.7 \it4 &&   ~851.88 \it3 &&$7^-$&&&              &         &   0.42 &&\\
 ~884.2 \it4 &&   ~883.9 \it2   &&$4^+$&$4^+$&&       0.51   &  1.27  &   4.09  & 0.8  & m1a\\
 ~923.3 \it5 &&   ~923.0 \it2   &&$6^+$&$6^+$&&       1.66  &   0.96  &   0.50  & 0.2  & m2d\\
 ~952.6 \it5 &&   ~951.88 \it5  &&$1^-$&$1^-$,&&       1.17  &   1.76  &   0.71  & 0.03 & m1a$^a$\\
             &&                 &&&or $(0^+)$&&        &         &         & 0.6  & sw.ij$^a$\\
         &&   ~955.1  \it2  &&$5^+$&  &&            &        &         &          &\\
 ~972.1 \it5 &&   ~971.72 \it5 &&$2^-$&$2^-$&&        0.22  &   0.69  &   0.37 & 5.0 & m2a\\
 1011.6 \it5 &&   1009.58 \it5 &&$2^+$&$2^+$&&        0.54   &  0.34  &  14.34 & 2.1 & sw.gg\\
             &&   1012.51 \it5 &&$3^-$&$3^-$&&               &        &        & 6.0 & sw.gg\\
 1040.0 \it7&&   1039.6 \it2 &&$6^+$&$6^+$&&        0.80   &  1.07  &   0.94 & 0.5 & m2a\\
 1052.0 \it7&&   1052.31 \it5 &&$3^+$&$3^+$&&        0.00  &   3.33  &   0.34 & 0.8 & m2a\\
 1065.9 \it8&&      &&&$4^-$&&        0.99   &  0.69   &  0.30 & 0.45 & m3b\\
 1079.4 \it8&&   1079.21 \it3 &&$2^-$&$2^-$&&        0.00   &  0.00   &  0.09 & 0.1 & m3b\\
 1108.7 \it5&&   1107.5 \it3 &&$4^+$&$4^+$&&        1.20   &  0.87  &   5.38 & 1.2 & m2a\\
      &&     1109.0 \it1 &&$5^-$&&&&& &&\\
      &&     1117.5 \it3 &&$8^+$&&&&& &&\\
 1125.6 \it5&&              &&&$(1^-)$,&&                4.54  &  1.06 &   2.10 & 0.1& m1a$^a$\\
            &&              &&&or $(0^+)$&&                   &       &        & 1.5& sw.jj$^a$\\
            &&   1127.76 \it5 &&$3^-$&&&                  &  &   & &  \\
    &&       1134.4 \it2 &&$7^+$&&&&& &&\\
 1148.0 \it9 &&             &&&     &&&&$<$0.2 &&\\
    &&      1176.1 \it3 &&$5^+$&&&&& &&\\
 1184.8 \it9 &&&&&&&&&$<$0.2 &&\\
    &&      1196.8 &&$(4^-)$&&&&& &&\\
 1241.2 \it9 && 1243.3 \it3 &&$8^+$&&&&&$<$0.2 &&\\
 1256.0 \it9 &&      1255.5 \it3 &&$6^+$&&&&&$<$0.2 &&\\
 1259.2 \it6&&              &&&$(3^-)$&&       0.97  &  0.34 &   0.34 &0.3&sw.gg\\
 1283.6 \it6&&          &&& $(5^-)$ &&          1.08  &    0.00 &     0.23 &0.07 & m3b\\
 1297.8 \it6&&   1297.1 \it1 &&$0^+$&$0^+$    &&         4.37 &    2.53 &    4.15 & 1.8 & sw.ig\\
 1322.3 \it5&&          &&&$(3^-)$&&          1.23  &   0.54  &   1.77 & 1.5 & sw.gg\\
 1337.2 \it5&&          &&&$4^+$&&             1.37  &   1.14   & 24.42 & 3.2 & sw.gg\\
          &&   1349.3 \it4   &&$7^+$&&& & & &&\\
 1359.5 \it7&&          &&&$(2^+)$&&          0.00  &   0.41  &   0.48 & 0.05 & sw.gg\\
 1376.6 \it7&&   1375.3 \it1 &&$1,2^+$&$1^+,5^-$&&      0.00   &  0.00  &   0.24 & 0.2/0.1 & m3b\\
 1401.5 \it5&&   1400.9 \it1 &&$2^+$&$2^+$  &&      0.49  &   0.38   &  3.18 & 0.3 & sw.gg\\
 1420.4 \it5&&          &&&$(3^+)$&&      0.33   &  0.70  &   4.74 & 1.0 & m2a\\
 1440.4  \it8&&              &&& $(3^+)$&&             &        &  $<$0.2      &     &     \\
 1447.9 \it5&&          &&&$0^+$  &&      10.37   &  6.86  &   1.70 & 0.05 & sw.gg\\
 1485.6 \it5&&   1485.6 \it1 &&&$4^+$  &&       1.44  &   0.93  &   4.59 & 0.6 & sw.gg\\
 1496.0 \it10 &&&&&&&&&$<$0.2 &&\\
 1507.4 \it5&&          &&&$4^+$&&       1.29  &   2.28 &    0.57 & 0.07 & m2a\\
 1524.8 \it5&&          &&&$2^+$&&       0.66  &  0.00 &   0.53 & 0.05 & sw.gg\\
 1566.2 \it6&&          &&&$(1^-)$&&       1.80  &   0.29  &   0.31 & 0.15 & m2b\\
 1574.5 \it6&&   1573.5 \it2 &&$1^{(-)},2^+$&$(2^-)$&&0.34  &   0.70  &   2.14 & 3.8 & m2a\\
 1584.7 \it6&&          &&&$(4^-,5^+)$&&  1.77  &   1.90  &   2.64 & 42/36 & m2a\\
 1590.2 \it5&&   1589.8 \it2 &&$0^+$&$0^+$ &&       8.48   &  6.04  &  11.66 & 0.28 & sw.gg\\
 1594.7 \it8&&          &&&$(1^-)$&&      1.11  &   0.58 &    2.22 & 1.5 & m3b\\
 1601.2 \it11&&         &&&$(3^-)$&&      1.46  &   1.14 &   0.74 & 0.8 & sw.gg\\
 1612.1 \it10&&         &&&$(4^-,5^+)$&&   1.26  &  2.25 &   0.36 & 8.3/1 & m2a\\
 1618.7 \it9 &&         &&&$(4^-,5^+)$&&   1.44  &  2.09   &   0.27  & 8.0/0.9 & m2a\\
 1630.1 \it7&&   1628 \it2  &&&$2^+$&&         1.43  &  0.45   &   4.49  & 0.6    & m1a\\
 1639.3 \it6&&   1638.5 \it2 &&$(2,0^+)$&$0^+$&&        5.68  &  4.64   &   8.32  &  0.2  &  sw.gg\\
 1653.2 \it11&&             &&&$(6^+)$&&           0.96  &   1.75  &   0.88  & 0.004   & m3a\\
 1668.2 \it7&&          &&&$4^+$&&     1.54   &  1.85  &   1.91  & 0.3 & sw.gg\\
 1679.1 \it7&&          &&&$2^+$&&         1.72  &   0.36  &   3.36  & 0.45   & m1a\\
 1683.3 \it7&&              &&&$(4^-)$&&       0.19  &   1.95  &   2.20  & 34     & m2a\\
 1694.9 \it7&&   1695.7 \it1    &&$1^{(-)},2^+$&$(4^+)$&&     1.09   &  1.41  &   3.57 & 0.35 & sw.gg\\
 1708.8 \it8&&             &&&$2^+$&&     0.22  &   0.18  &   0.38 & 0.03 & sw.gg\\
 1723.5 \it7&&         &&& $(4^+)$ &&      0.55  &   1.35  &   2.31 & 0.3 & m1a\\
 1745.3 \it8&&   1744.9 \it1 &&&$0^+$    &&     2.80  &   2.59  &   1.18 & 0.10 & sw.jg\\
 1750.7 \it8&&          &&&$(3^-)$&&       1.76  &   0.64  &   0.99 & 0.8 & sw.gg\\
 1762.3 \it8&&              &&&$(4^+)$&&       0.73  &  0.43 &   0.89 & 0.2 & m1a\\
 1769.6 \it8&&   1770.7 \it1 &&$1,2^+$&$(4^+)$&&       0.52 &   0.69 &   0.89 & 0.15 & sw.gg\\
 1774.1 \it9 &&   1775.2 \it1  &&$1,2^+$&&& &&$<$0.2&\\
      &&   1789.4 \it5 &&$1^{(-)},2^+$&&&&& &&\\
 1793.1 \it6&&          &&&$(5^-)$&&        1.19 &   1.19 &   2.43 & 3.5 & sw.gg\\
 1802.5 \it6&&              &&&$0^+$  &&        8.04  &   5.38 &   2.38 & 0.05 & sw.gg\\
      &&   1810.7 \it1 &&$1,2^+$&&&&& &&\\
 1812.0 \it8&&             &&&$4^+$&&       1.71 &   0.96 &   1.26 & 0.13 & sw.gg\\
 1824.9 \it7&&              &&&$(6^+)$&&        0.45 &   0.96 &   1.15 & 0.15 & sw.gg\\
 1840.0 \it8&&   1839.6 \it2 &&$1^{(-)},2^+$&$2^+$&&        1.74 &   0.70 &   2.37 & 0.25 & m1a\\
 1851.4 \it7&&   1849.6 \it1 &&$(2^+)$&$(3^-)$&&     1.07 &   0.93 &   1.73 &1.5 & sw.gg\\
 1859.3 \it7&&   1858.2 \it6 &&&$(3^-)$&&       1.41 &   1.05 &   1.74  & 1.6& sw.gg\\
 1868.9 \it7&&          &&&$(0^+)$&&        1.99 &   1.60 &   1.73 &  0.03 & sw.gg$^b$\\
                          &&&&&$+(6^+)$&&                 &        &        &  0.26 & m2d$^b$\\
 1887.0 \it9&&          &&&$(2^+)$&&          0.84  &  0.68 &   0.76 & 0.06 & sw.gg\\
      &&   1902.7 \it1  &&$1,2^+$&&&&&&&\\
 1910.0 \it9&&          &&&$(6^+)$&&        0.51  &   1.13 &  1.84 &  0.25 & sw.gg\\
 1914.7 \it9&&              &&&$(1^-)$&&        0.91  &   0.57 &   1.18 & 0.8 & m2a\\
 1926.0 \it7&&              &&&$4^+$&&              2.18  &   0.90 &   1.71 & 0.6 & m2a\\
 1931.1 \it8&&              &&&$(1^-)$&&        0.81  &   0.50 &   0.62 & 0.5 & m2a\\
 1939.8 \it11&&             &&&$(1^-,1^+)$&&        0.36  &   0.68  &   0.98 & 0.6 & m2a\\
 1947.0 \it6&&              &&&$4^+$&&       1.18  &  0.89 &   3.81 & 0.4 & m1a\\
      &&   1949.8 \it1 &&$1,2^+$&&&&& &&\\
 1956.4 \it6&&          &&&$2^+$&&        0.39  &  0.47 &   6.96 & 0.35 & sw.gg\\
 1967.1 \it7&&   1966.9 \it2 &&$1,2^+$&$2^+$&&        0.27  &  0.66 &   3.76 & 0.18 & sw.gg\\
 1972.0 \it9&&   1973.4 \it2 &&$(1^+,2^+)$&$2^+$&&      0.58  &  0.39 &   0.94 & 0.05 & sw.gg\\
 1985.4 \it8&&          &&&$(5^-)$&&    0.00   & 0.52 &   0.51 & 0.8 & sw.gg\\
 2001.6 \it8&&   2000.9 \it1 &&$1,2^+$&$(3^-)$&&       1.28  &  0.83 &   1.15 & 1.1 & sw.gg\\
 2010.3 \it6&&   2010.1 \it2 &&$1,2^+$&$2^+$ &&       0.26  &  0.38 &   5.69 & 0.3 & sw.gg\\
 2017.3 \it7&&          &&&$(3^-)$&&          1.03  &  0.73 &   1.75 & 1.6 & sw.gg\\
 2025.6 \it6&&   2024.7 \it2 &&$1^+,2^+$&$2^+$ &&       0.28  &  0.39 &   3.84 & 0.2 & sw.gg\\
 2032.8 \it7&&          &&&$4^+$&&    1.07 &   1.13  &  5.69 & 0.5 & sw.gg\\
 2039.1 \it7&&              &&&$4^+$&&        2.24 &   1.85 &   4.51 & 0.4 & sw.gg\\
 2048.7 \it7&&              &&&$(4^+)$&&          1.00 &   1.19 &   1.59 & 0.15 & sw.gg\\
 2060.9 \it12&&             &&&$(3^-)$&&              2.05  &  0.55 &   0.87 & 0.02 & m3a\\
 2073.2 \it8&&          &&&$(8^+)$&&              0.00  &  1.82 &   1.40 & 0.4 & sw.gg\\
 2074.9 \it8&&      &&&$(4^+)$&&                  0.85 &   0.51 &   1.16& 0.2 & m1a\\
 2085.9 \it8&&      &&&$(4^+)$&&                  1.75  &  1.26 &   1.60& 0.15 & sw.gg\\
 2093.9 \it7&&      &&&$0^+$  &&             4.44 &   2.17 &   6.93& 13 & sw.ii\\
 2102.0 \it7&&          &&&$4^+$&&            2.30  &  1.10 &   1.26& 0.15 & sw.gg\\
 2118.4 \it6&&      &&& $4^+$ &&                2.32 &   1.14 &   7.70& 0.90 & sw.gg\\
            &&  2122.8 \it1 &&$1,2^+$&&&               &        &       &      & \\
 2130.7 \it7&&                &&&$2^+$&&0.51  &  0.76 &   5.23& 0.2 & sw.gg\\
        &&  2133.2 \it2 &&& &&&&&&\\
 2137.9 \it7&&      &&&$2^+$&&            0.51  &  0.50 &   4.87& 0.2 & sw.gg\\
 2150.5 \it6&&      &&&$0^+$  &&              10.53 &   3.87 &   6.57& 25 & sw.ii\\
         &&   2151.8 \it2 &&$1,2^+$&&& &  &&&   \\
 2168.8  \it7&&     &&&$(4^+)$&&                  1.12   &  1.37 &    2.31 & 0.2  & sw.gg\\
 2175.1 \it6 &&     &&&$0^+$       &&          21.75   &  8.97  &  11.93 & 42 & sw.ii\\
 2181.7 \it7&&      &&&$(4^+)$&&              1.02  &   2.07   &  4.22 & 0.35 & sw.gg\\
 2187.1 \it6&&      &&&$2^+$&&            0.39  &   0.57  &   8.14 & 0.30 & sw.gg\\
 2194.8 \it8 &&         &&&$(6^+)$    &&             &     &        0.59 & 0.25 & m2a\\
 2205.4 \it10&&     &&&$2^+$&&                  0.49   &  0.47  &   2.67 & 0.10 &sw.gg\\
 2207.8 \it8&&      &&&$(4^+)$&&              1.99  &   2.40  &   2.74 &  0.2 & sw.gg\\
 2216.0 \it7 &&     &&&$(4^+)$&&              1.07   &  0.44   &  1.62 & 0.4 & m2a\\

 2226.0 \it6 &&     &&&$2^+$&&            0.56   &  0.54   &  9.84 & 0.5 & sw.gg\\
 2241.0 \it7 &&     &&&$2^+$&&              0.29   &  0.49   &  1.13 & 0.06 & sw.gg\\
 2249.9 \it7 &&     &&&$(6^+)$&&         0.46   &  1.22  &   1.54 & 0.2 & sw.gg\\
 2255.3 \it7&&      &&&$4^+$    &&          1.81   &  1.41  &   3.79 & 0.4 & sw.gg\\
 2268.9 \it6&&      &&&$0^+$ &&          15.37  &   6.16  &  14.86 & 56 & sw.ii\\
 2276.0 \it8&&      &&&$(4^+)$&&              0.00  &   2.02 &    1.59 & 0.15 & sw.gg\\

 2282.1 \it10&& 2282.8 \it5 &&$1,2^+$&&& & & &&\\
 2295.9 \it8&&      &&&$4^+$&&              3.08  &   1.78  &   2.20 & 0.25 & sw.gg\\
    &&  2298.6 \it3 &&$1,2^+$&&& & & &&\\
 2305.4 \it7&&      &&&$2^+$&&            0.52  &   0.53  &  13.62 & 0.7 & sw.gg\\
 2311.2 \it8&&      &&&$(4^+)$&&      0.41  &   1.20  &   2.33 & 0.2 & sw.gg\\
    &&  2314.3 \it2 &&$1,2^+$&&& & & &&\\
 2317.7 \it7&&      &&&$4^+$&&                2.75  &   1.27  &   4.95 & 0.5 & sw.gg\\
 2329.6 \it7&&      &&&$2^+$&&            0.37  &   0.34  &   2.65 & 0.1 & sw.gg\\
 2337.1 \it8&&      &&&$(5^-)$&&              0.90   &  0.87   &  1.28 & 2.0 & sw.gg\\
 2354.8 \it10&&     &&&$(6^+)$&&              2.12  &   2.20   &  0.60 & 1.6 & m2a\\
 2368.5 \it7&&  2368.9 \it2 &&&$(0^+)$&&          2.75  &   3.79   &  1.70 & 0.23 & sw.jg\\
 2383.8 \it8&&      &&&$(4^+)$&&              1.02  &   0.87  &   4.54 & 0.3 & m2a\\
 2388.4 \it10&&     &&&       &&                    &         &  1.32&&\\
 2395.2 \it7&&      &&&$0^+$ &&             3.95 &   2.14  &  0.94& 2.8 & sw.ii\\
 2402.0 \it8&&      &&&$(6^+)$&&              0.50 &   0.70  &   0.89& 0.4 & m2a\\
 2411.6 \it7&&      &&&$2^+$&&            0.30  &  0.60 &   6.39& 0.3 & sw.gg\\
 2422.7 \it7&&      &&&$(4^+)$ &&             3.20 &   1.87 &  8.3& 0.8 & sw.gg$^b$\\
            &&          &&&$+(0^+)$ &&                  &        & 5.2& 7.5 & sw.ii$^b$\\
 2426.4 \it9&&      &&&$(0^+)$ &&         3.11 &   2.30 &   3.50& 6.7 & sw.ii\\
 2436.6 \it9&&      &&&$2^+$&&            0.33 &   0.49 &   2.20& 0.10 & sw.gg\\
 2442.5 \it8&&      &&&$2^+$&&            0.65 &   0.68 &   3.73& 0.15 & sw.gg\\
 2449.2 \it2&&      &&&$(3^-)$&&              1.34 &   0.71 &   1.65&1.6 & sw.gg\\
 2461.0 \it7&&      &&&$2^+$&&            0.41 &   0.62 &   8.08& 0.4 & sw.gg\\
 2467.2 \it7&&      &&&$2^+,3^-$&&            0.49 &   0.79 &   3.55& 0.15 & sw.gg\\
 2474.3 \it8&&      &&&$2^+$&&              0.15 &   0.50  &  5.20& 0.25 & sw.gg\\
 2478.5 \it8&&      &&&$4^+$&&              2.20 &   1.00  &  5.00& 0.5 & sw.gg\\
 2481.3 \it12&&     &&&$(6^+)$&&              0.03 &   0.16 &   1.21& 0.5 & m2a\\
 2493.8 \it7&&      &&&$0^+$   &&           5.62 &   3.92 &   3.40& 1.4 & sw.ig\\
 2501.1 \it7&&      &&&$4^+$&&            1.43 &   1.41 &   4.70& 0.6 & sw.gg\\
 2508.3 \it7&&      &&&$ $&&               0.00 &   1.68 &   0.76&&\\
 2519.3 \it7&&      &&&$(6^+)$&&              0.00 &   1.13 &   1.43& 0.5 & m2a\\
 2528.1 \it7&&      &&&$0^+$   &&            11.96 &   6.76 &  12.36& 5.6 & sw.ig\\
 2536.9 \it7&&      &&&$4^+$&&            1.36 &   1.43 &   6.07& 0.6 & sw.gg\\
 2549.8 \it11&&     &&&$0^+$   &&             4.32 &   2.60 &   2.75& 1.2 & sw.ig\\
 2556.2 \it8&&      &&&$(4^+)$&&              0.85  &  1.10 &   4.05& 0.75 & m1a\\
 2562.9 \it9&&      &&&$(4^+)$&&              0.00 &   0.83  &  1.56& 0.2 & sw.gg\\
 2573.2 \it7&&      &&&$(6^+)$&&              0.00 &   1.18 &   1.33& 0.5 & m2a\\
 2589.1 \it7 &&     &&&$2^+$&&            0.49  &  0.64 &   3.92& 0.15 & sw.gg\\
 2596.4 \it8&&      &&&$(0^+)$&&             31.20 &  10.35 &   2.50& 5.1 & sw.ii\\
 2601.3 \it7 &&     &&&$(4^+)$&&              1.34 &   0.93 &  14.48& 2.8 & m2a \\
 2616.0 \it7 &&     &&&$2^+$&&            0.55 &   0.61 &   2.76& 0.4 & sw.gg\\
 2625.9 \it7&&      &&&$2^+$&&            0.37 &   0.40 &   4.41& 0.15 & sw.gg\\
 2640.0 \it8&&      &&&$4^+$&&            1.83 &   1.55 &   7.41& 0.7 & sw.gg\\
 2660.9 \it7&&      &&&$4^+$&&              2.92  &  2.29 &   4.24& 0.3 & sw.gg\\
 2666.4 \it7&&      &&&$(2^+)$&&              0.34 &   0.58 &  26.48& 1.5 & sw.gg\\
 2671.6 \it7&&      &&&$4^+$&&            1.77 &   1.07  & 11.82& 1.2 & sw.gg\\
 2679.2 \it8&&      &&&$2^+$&&            0.39 &   0.30 &   3.77& 0.15 & sw.gg\\
 2694.9 \it7&&      &&&$2^+$&&              0.49 &   0.43 &   1.25& 0.06 & sw.gg\\
 2706.5 \it7&&      &&&$2^+$&&              0.79 &   0.53 &   6.14& 0.2 & sw.gg\\
 2712.9 \it8&&      &&&$(6^+)$&&              0.74 &   1.17 &   2.84& 1.0 & m2a\\
 2726.6 \it7&&      &&&$2^+$&&            1.02 &   0.87  &  2.28& 0.1 &sw.gg\\
 2740.6 \it7&&      &&&$2^+$&&            0.33 &   0.41 &   5.13& 0.15 & sw.gg\\
 2746.2 \it7&&      &&&$4^+$&&            3.35 &   1.84 &   2.77& 0.3 & sw.gg\\
 2754.2 \it10&&     &&&$(6^+)$&&              1.47 &   1.19  &  1.26& 0.2 & sw.gg\\
 2764.9 \it7&&      &&&$2^+$&&            0.66  &  0.70 &   8.03& 0.3 & sw.gg\\
 2777.3 \it7&&      &&&$2^+$&&            0.78 &   0.51 &   7.89& 0.3 & sw.gg\\
 2791.5 \it7&&      &&&$4^+$&&            1.77 &   1.09  & 11.49& 1.2 & sw.gg\\
 2799.7 \it8&&      &&&$2^+$&&            0.25 &   0.51  &  3.01& 0.1 & sw.gg\\
 2808.1 \it7&&      &&&$0^+$  &&            7.72 &   3.30 &   4.86& 8.5 & sw.ii\\
 2824.4 \it10&&         &&&$4^+$&&                2.63 &   1.35 &   3.03 & 0.4 & sw.gg\\
 2834.0 \it10&&         &&&$2^+$&&              0.18 &   0.69 &   2.58 & 0.2 & sw.gg\\
 2841.3 \it7&&          &&&$(2^+)$&&              0.54 &   0.97 &   4.26 & 0.3 & sw.gg\\
 2855.9 \it7&&          &&&$2^+$&&            0.51 &   0.82 &   4.04 & 0.3 & sw.gg\\
 2862.9 \it7&&          &&&$2^+$&&            0.58 &   0.74 &   6.25 & 0.4 & sw.gg\\
 2870.6 \it10&&         &&&$(3^-)$&&              1.50  &  0.93 &   1.26 & 1.2 & sw.gg \\
 2879.7 \it7 &&         &&&$2^+$&&            0.76 &   0.62 &   5.49 & 0.5 & sw.gg\\
 2886.1 \it10&&         &&&$(1^-)$&&              1.71 &   0.49 &   0.99 & 0.8 & m2a\\
 2896.1 \it7 &&         &&&$2^+$&&            0.32  &  0.51 &   5.52 & 0.5 & sw.gg\\
 2906.4 \it8 &&         &&&$(3^-)$&&              0.75 &   1.07 &   3.42 & 2.6 & sw.gg\\
 2913.6 \it15&&         &&&$(4^+)$&&              0.51 &   0.37 &   1.60 & 0.12 & sw.gg\\
 2923.7 \it9 &&         &&&$2^+$&&            0.51 &   0.85 &   5.87 & 0.4 & sw.gg\\
 2930.6 \it7 &&         &&&$2^+$&&            0.49 &   0.50 &   6.31 & 0.26 & sw.gg\\
 2940.6 \it7 &&         &&&$2^+$&&            0.37 &   0.53  &   3.84 & 0.3 & sw.gg\\
 2950.5 \it8 &&         &&&$(6^+)$&&              0.20 &   1.41  &   0.99 & 0.5 & m2a\\
 2987.9 \it10&&         &&&$(6^+)$&&              0.49 &   0.96  &   3.75 & 0.6 & sw.gg\\
 2999.0 \it7 &&         &&&$2^+$&&            0.46 &   0.71  &   7.66 & 0.6 & sw.gg\\
 3009.9 \it8 &&         &&&$2^+$&&              0.34 &   0.67  &   2.78 & 0.2 & sw.gg\\
 3020.6 \it8 &&         &&&$2^+$&&            0.38 &   0.52  &   3.72 & 0.3 & sw.gg\\
 3030.3 \it9 &&         &&&$2^+$&&              0.40 &   0.95  &   3.83 & 0.3 & sw.gg\\
 3043.0 \it7 &&         &&&$2^+$&&            0.46 &   0.61  &   5.91 & 0.4 & sw.gg\\
 3052.4 \it9 &&         &&&$(3^+)$&&              0.51 &   1.08  &   3.53 & 4.2 & m2a\\
 3064.3 \it15&&         &&&$(2^+)$&&              0.48 &   0.64  &   2.51 & 0.1 & sw.gg\\
 3072.6 \it8 &&         &&&$(6^+)$&&              0.64 &   1.20  &   3.82 & 0.5 & sw.gg\\

 3083.8 \it7 &&         &&&$2^+$&&            0.36 &   0.64  &   5.46 & 0.35 & sw.gg\\
 3100.9 \it7 &&         &&&$2^+$&&            0.41 &   0.58  &   6.07 & 0.40 & sw.gg\\
 3113.9 \it12&&         &&&$(\leq 4)$       &&      1.04 &   1.48  &   2.57 &&\\
 3124.7 \it8 &&         &&&$(4^+)$      &&      0.89 &   1.15  &   6.01 & 1.8 & sw.gg\\
 3135.9 \it10&&         &&&$(\leq 4)$       &&      1.42 &   1.10  &   4.52 & 0.7& sw.gg\\
 3147.4 \it8 &&         &&&$(\leq 4)$        &&     1.06 &   0.81  &   4.65 &&\\
 3162.0 \it7 &&             &&&$2^+$        &&      0.41 &   0.79  &   3.44 & 0.25 & sw.gg\\
 3173.6 \it8 &&         &&&$2^+$        &&      0.27  &  0.63  &   3.29 & 0.25 & sw.gg\\
 3186.1 \it7 &&         &&&$(6^+)$      &&      0.46 &   0.82  &   2.54 & 0.4 & sw.gg\\
 3198.4 \it7 &&         &&&$2^+$      &&      0.63 &   0.72  &   3.56 & 0.1&sw.gg\\
 3212.2 \it7 &&         &&&$2^+$      &&      0.50 &   0.73  &   2.75 &0.08&sw.gg\\
 3223.1 \it7 &&         &&&$2^+$      &&      0.54 &   0.79  &   3.14 &0.08&sw.gg\\
 3234.0 \it7 &&         &&&     &&      0.98 &   0.89  &   4.72 &&\\
 3248.6 \it7 &&         &&&$2^+$      &&      0.54 &   0.72  &   4.46 &0.12&sw.gg\\
 3258.8 \it8 &&         &&&     &&      1.38  &  1.33  &   5.09 &&\\
 3269.9 \it12&&         &&&$(2^+)$      &&      0.85 &   0.83  &   4.29 &0.12&sw.gg\\
 \hline
\end{longtable*}
\footnotesize \noindent
 $^{a)}$  For levels at 952.6 and 1125.6
keV the angular distributions can be fitted by the DWBA
calculations as spin 1$^-$ or  spin 0$^+$.\\
$^{b)}$  For the levels at 1868.9 and 2422.7 keV the angular
distributions are fitted by the DWBA calculations as
doublet lines.\\
\normalsize

\subsection{\label{sec:DWBA} DWBA analysis}

The spins of the excited states in the final nucleus $^{230}$Th
were assigned via an analysis of the angular distributions of
tritons from the (p,t) reaction. In  a previous publication
\cite{Wir04} the angular distributions for $0^+$ excitations were
demonstrated to have a steeply rising cross
 section at very small reaction  angles and a sharp minimum at
a detection angle of about 14$^\circ$. The angular distribution
for the $0^+$ ground  state of $^{230}$Th has such a shape. This
pronounced feature helped us to identify these states in
complicated and dense spectra without fitting experimental angular
distributions. No complication of the angular
 distributions was expected, since the excitation of  $0^+$ states
predominantly is a one-step process.  This is not the case for the
excitation of  states with other spins, where multi-step processes
could play a very important role.

%%%%%%%%%%%%%%%%%%%%%%%%%%%%%%%%%%%%%%%%%%%%%%%%%%%%%%%%%%%%%%%%%%%
\begin{figure}
\begin{center}
%fig3
\epsfig{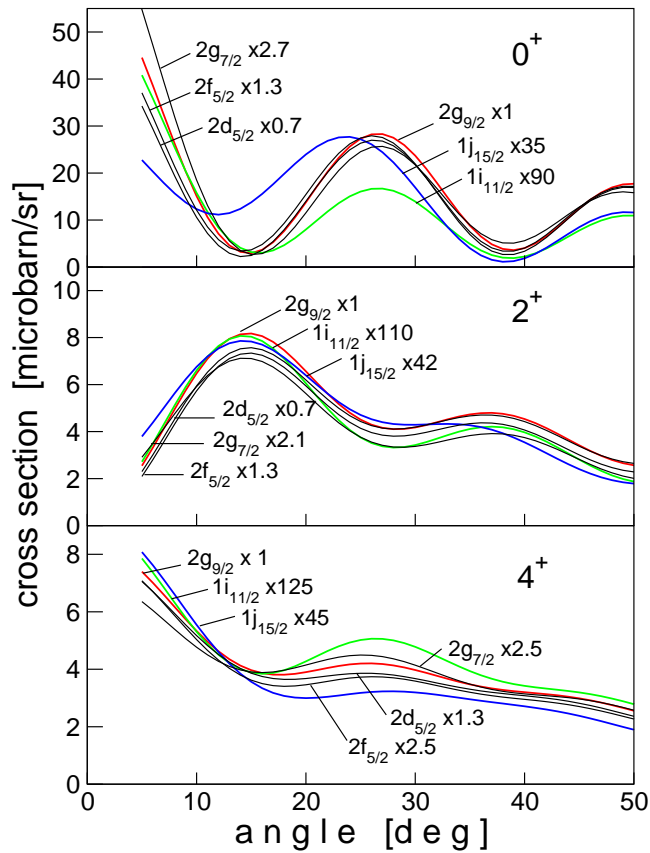}
\caption{\label{fig:angl_distr_cal} CHUCK3 one-step DWBA
calculations of the angular distributions for different $(j)^2$
transfer configurations. The lines are marked with the $(j)^2$
transfer configuration and a scaling factor introduced to allow
comparison with the $(2g_{9/2})^2$ transfer configuration.}
\end{center}
\end{figure}

%%%%%%%%%%%%%%%%%%%%%%%%%%%%%%%%%%%%%%%%%%%
%%%%%%%%%%%%%%%%%%%%%%%%%%%%%%%%%%%%%%%%%%%%%%%%%%%%%%%%%%%%%%%%%%%

\newcolumntype{d}{D{.}{.}{2}}
\begin{table}
\caption {\label{tab:IBM_potent} Optical potential parameters used
in the DWBA calculations.}
 \begin{ruledtabular}
 \begin{tabular}{clddcd}
 \smallskip\\
 \multicolumn{2}{c}{Parameters}
   & \multicolumn{1}{c}{p}
   & \multicolumn{1}{c}{t\footnotemark[1]}
   & \multicolumn{1}{c}{n}
   & \multicolumn{1}{c}{t\footnotemark[2]}\\
   \smallskip\\
 \hline
\smallskip\\
$V_r$   &(MeV) &57.10 &166.70 &         &159.0\\
$4W_D$  &(MeV) &32.46 &       &         &         \\
$W_0$   &(MeV) &2.80  &10.28  &         &9.24      \\
$4V_{so}$&(MeV)&24.80 &       &$\lambda\,=\,25$& \\
$r_r$   &(fm)  &1.17  &1.16   &1.17     &1.16     \\
$r_D$   &(fm)  &1.32  &       &         & \\
$r_0$   &(fm)  &1.32  &1.50   &         &1.50     \\
$r_{so} $&(fm) &1.01  &       &         &         \\
$R_c$   &(fm)  &1.30  &1.30   &         &1.25     \\
$a_r$   &(fm)  &0.75  &0.75   &0.75     &0.75     \\
$a_D$   &(fm)  &0.51  &       &         &         \\
$a_{0}$ &(fm)  &0.51  &0.82   &         &0.82 \\
$a_{so}$&(fm)  &0.75  &       &         &        \\
$nlc$   &      &0.85  &0.25   &         &0.25 \\
   \end{tabular}
\end{ruledtabular}
\footnotetext[1]{According to \protect \cite{Fly69}.}
\footnotetext[2]{According to \protect \cite{Bec71}.}
\end{table}

%%%%%%%%%%%%%%%%%%%%%%%%%%%%%%%%%%%%%%%%%%%%%%%%%%%%%%%%%%%%
\begin{figure*}
\begin{center}
\epsfig{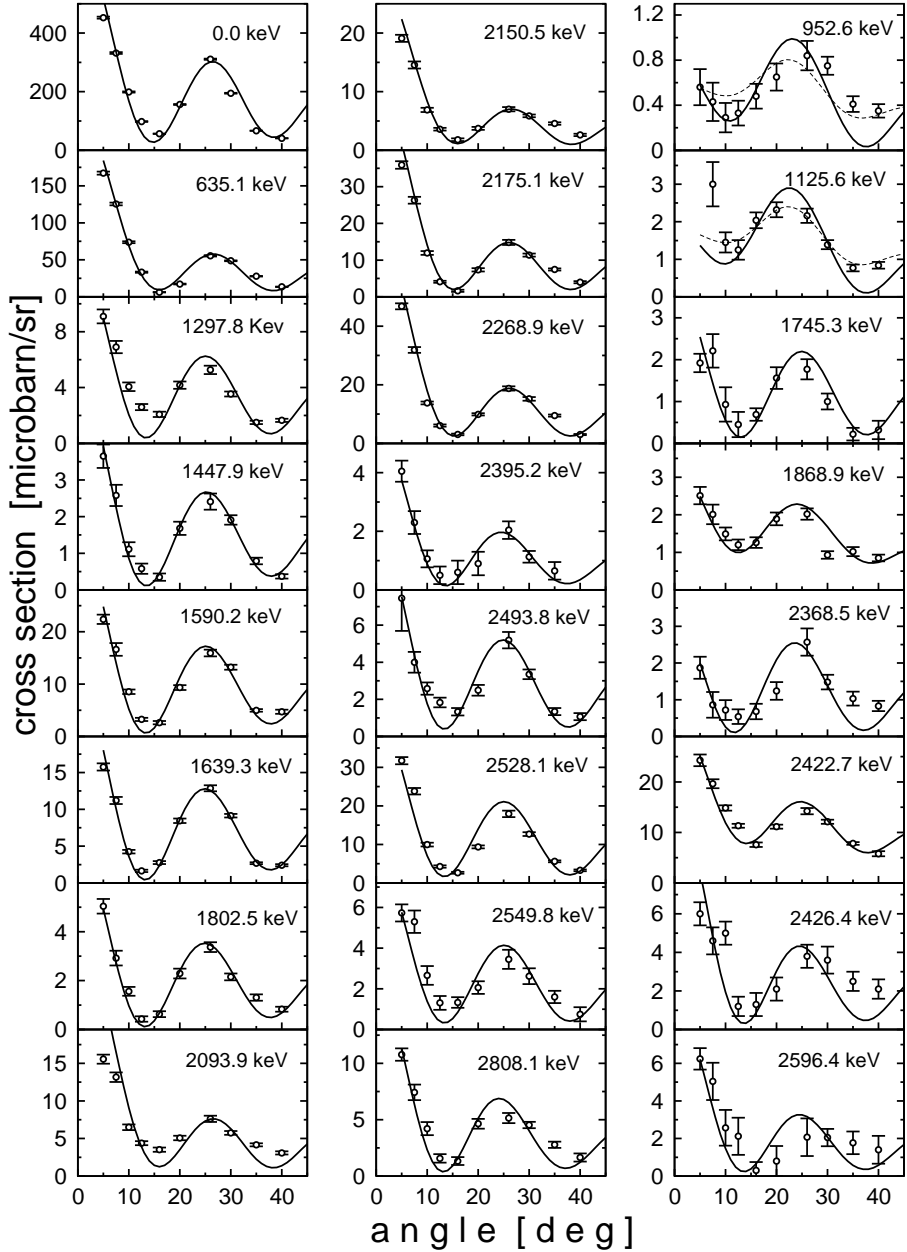}
    \caption{\label{fig:angl_distr_0+}
    Angular distributions of assigned $0^+$ states in $^{230}$Th and
     their fit with  CHUCK3 one-step calculations. The  $(ij)$ transfer
     configurations used in the calculations for the best fit are
    given in Table~\ref{tab:expEI}. The first two columns on the left
    correspond to firm assignments and the column on the right
    to tentative assignments. }
\end{center}
\end{figure*}

%%%%%%%%%%%%%%%%%%%%%%%%%%%%%%%%%%%%%%%%%%%%%%%%%%%%%%%%

The identification of other states is possible by fitting the
experimental
 angular distributions with those calculated in the distorted-wave
Born approximation (DWBA). A problem arising in such calculations
is that we have no prior knowledge of the microscopic structure of
these states. We can assume, however, that  the overall shape of
the angular distribution of the cross section  is rather
 independent of the specific structure of the individual states,
since the wave function of the outgoing tritons is restricted to
the nuclear exterior and therefore to the tails of the triton form
factors. To verify this assumption  DWBA calculations of angular
distributions for different $(j)^2$  transfer configurations to
states with different spins were carried out. The coupled channel
approximation (CHUCK3 code of Kunz  \cite{Kun}) was used in these
calculations.

The shape of the calculated angular distributions depends strongly
on the chosen potential parameters. We used parameters suggested
by Becchetti and Greenlees \cite{Bec69} for protons and by Flynn
et al. \cite{Fly69} for tritons. These
 parameters have been tested via their description of angular distributions
 for the ground states of $^{228}$Th,  $^{230}$Th and $^{232}$U
 \cite{Wir04}. They are listed in Table~\ref{tab:IBM_potent}.
 Minor changes of
the parameters for tritons were needed only for some $3^-$ states,
particularly for the state at 571.7 keV. For these states the
triton potential parameters suggested by Becchetti and Greenlees
\cite{Bec71} were used (the last column in
Table~\ref{tab:IBM_potent}). For each state the binding energies
of the two neutrons are calculated to match the outgoing triton
energies. The corrections to the reaction
 energy are introduced depending on the excitation energy.
The best reproduction of the angular distribution for the ground
state was obtained for the  transfer of the $(2g_{9/2})^2$
configuration in the one-step process.
 This orbital is close to the  Fermi surface and was considered as
the most probable in the transfer process.  Other transfer
 configurations that might be of importance are $(1i_{11/2})^2$ and
 $(1j_{15/2})^2$,
since these orbitals are also near the Fermi surface. All other
configurations for the orbitals in the vicinity of the Fermi
surface were tested in the one-step calculations. As one can see
in Fig.~\ref{fig:angl_distr_cal}, the shape of the angular
distributions depends to some degree on the transfer
configuration, the most pronounced being found for the 0$^+$
states. However, the main features of the angular distribution
shapes for 2$^+$ and 4$^+$ states are not dependent on the
transfer configurations. Therefore the $(2g_{9/2})^2$
configuration
 was used in the calculations for the majority of  excited states.
For some $0^+$ excitations the configurations
 $(1i_{11/2})^2$ and $(1j_{15/2})^2$ alone or in combination with $(2g_{9/2})^2$
were needed  to obtain better reproduction of the experimental
angular distributions.

%%%%%%%%%%%%%%%%%%%%%%%%%%%%%%%%%%%%%%%%%%%%%%%%%%%%%%%%%%%%%%%
\begin{figure*}
\begin{center}
%fig5
\epsfig{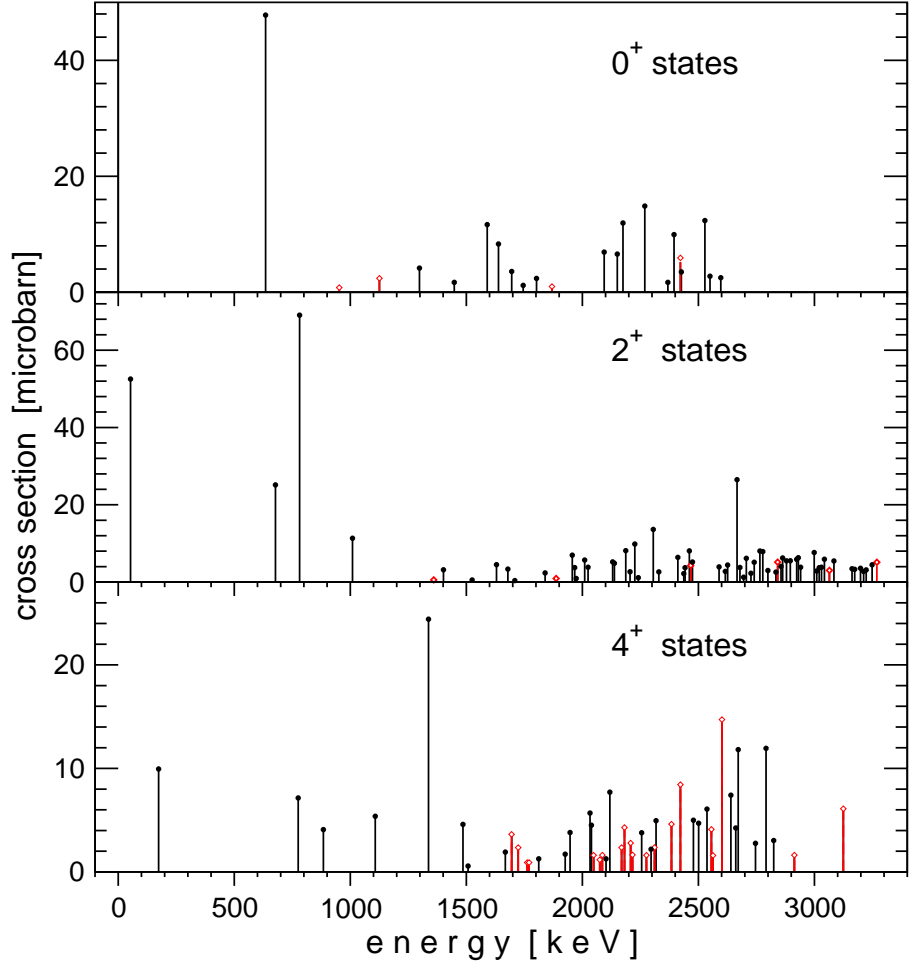}
    \caption{\label{fig:strength} Experimental  distribution of
    the (p,t) strength integrated in the angle region 0$^\circ$  - 45$^\circ$
    for 0$^+$, 2$^+$ and 4$^+$ states in $^{230}$Th.
    The levels  identified reliably are indicated by filled circles and
    those identified tentatively  are indicated by open diamonds.}
\end{center}
\end{figure*}
%%%%%%%%%%%%%%%%%%%%%%%%%%%%%%%%%%%%%%%%%%%%%%%%%%%%%%%%%%%%%%%

\begin{figure}
\begin{center}
\epsfig{file=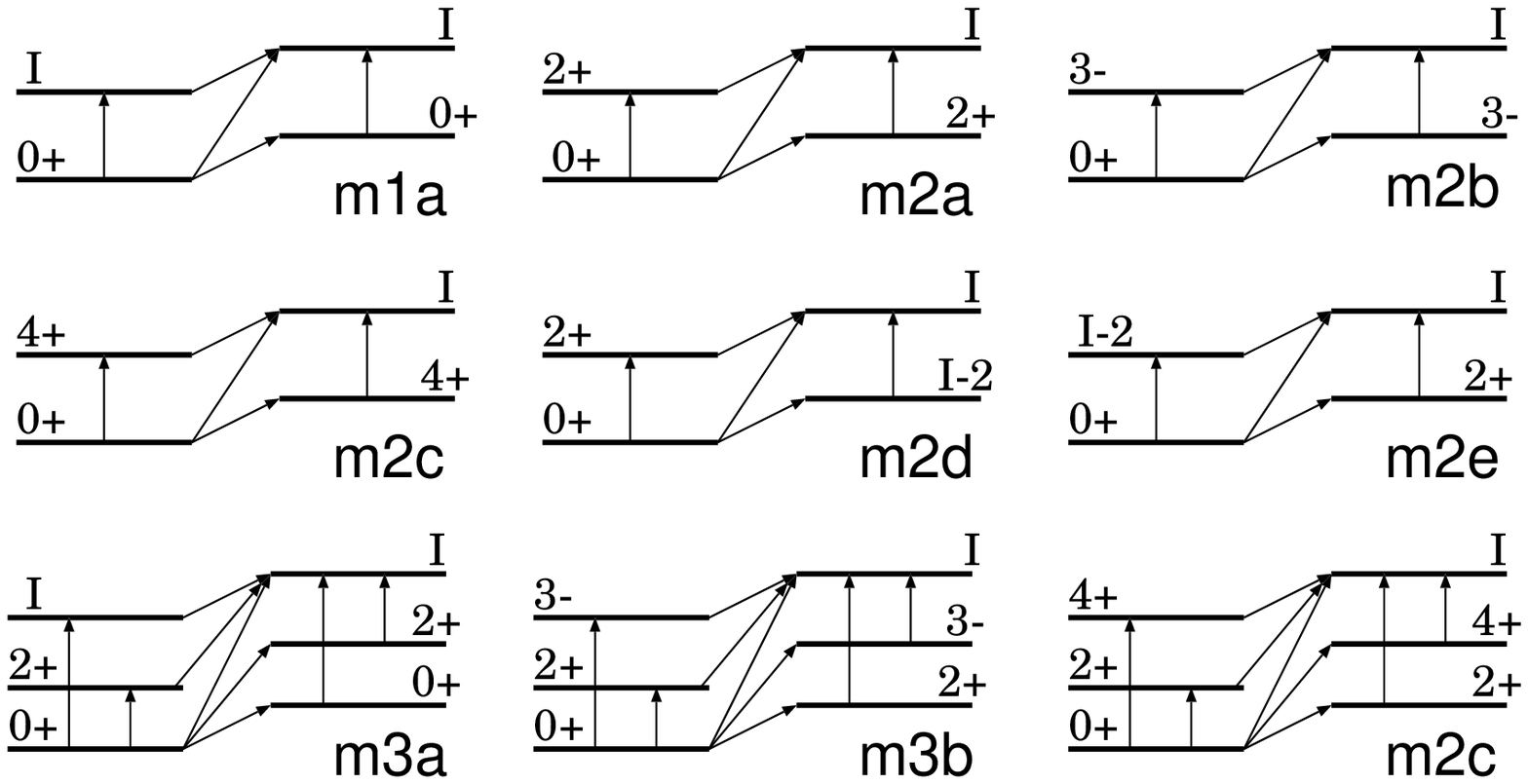, width=8.6cm,angle=0}
    \caption{\label{fig:schemes} Schemes of the CHUCK3 multi-step
    calculations tested with spin assignments of  excited states
    in $^{230}$Th (see Table~\ref{tab:expEI}).}
\end{center}
\end{figure}

%%%%%%%%%%%%%%%%%%%%%%%%%%%%%%%%%%%%%%%%%%%%%%%%%%%%%%%%%%%%%%%%%%%%%

Results of  fitting the angular distributions for the states
assigned as $0^+$ excitations are shown in
Fig.~\ref{fig:angl_distr_0+}. The angular distributions  in the
two first columns measured with good
 statistical accuracy are believed to give firm $0^+$ assignments.
 Three possible transfer configurations $(2g_{9/2})^2$,
$(1i_{11/2})^2$ and $(1j_{15/2})^2$  have been used to get the
best fit to the experimental data. These configurations are listed
in the last column in Table~\ref{tab:expEI}. The assignments to
the states of 1745.3, 2368.5 2426.4 and 2596.4 keV are considered
as relatively firm, because the shape of the angular distribution
is fitted perfectly by the calculations although there is a
limited statistical accuracy of the experimental data. The angular
distribution for the transition at the excitation energy 1868.9
keV is fitted as a sum of two angular distributions for transfer
to a $0^+$ and a $6^+$ state (doublet line). A similar situation
is assumed  for the energy 2422.7 keV, i.e. a superposition of two
angular distributions for $0^+$ and $4^+$ states.  In both  cases
assignments are tentative.

%%%%%%%%%%%%%%%%%%%%%%%%%%%%%%%%%%%%%%%%%%%%%%%%%%%%%%%%%%%%%%%%%%%%%%%
\begin{figure*}
\begin{center}
\epsfig{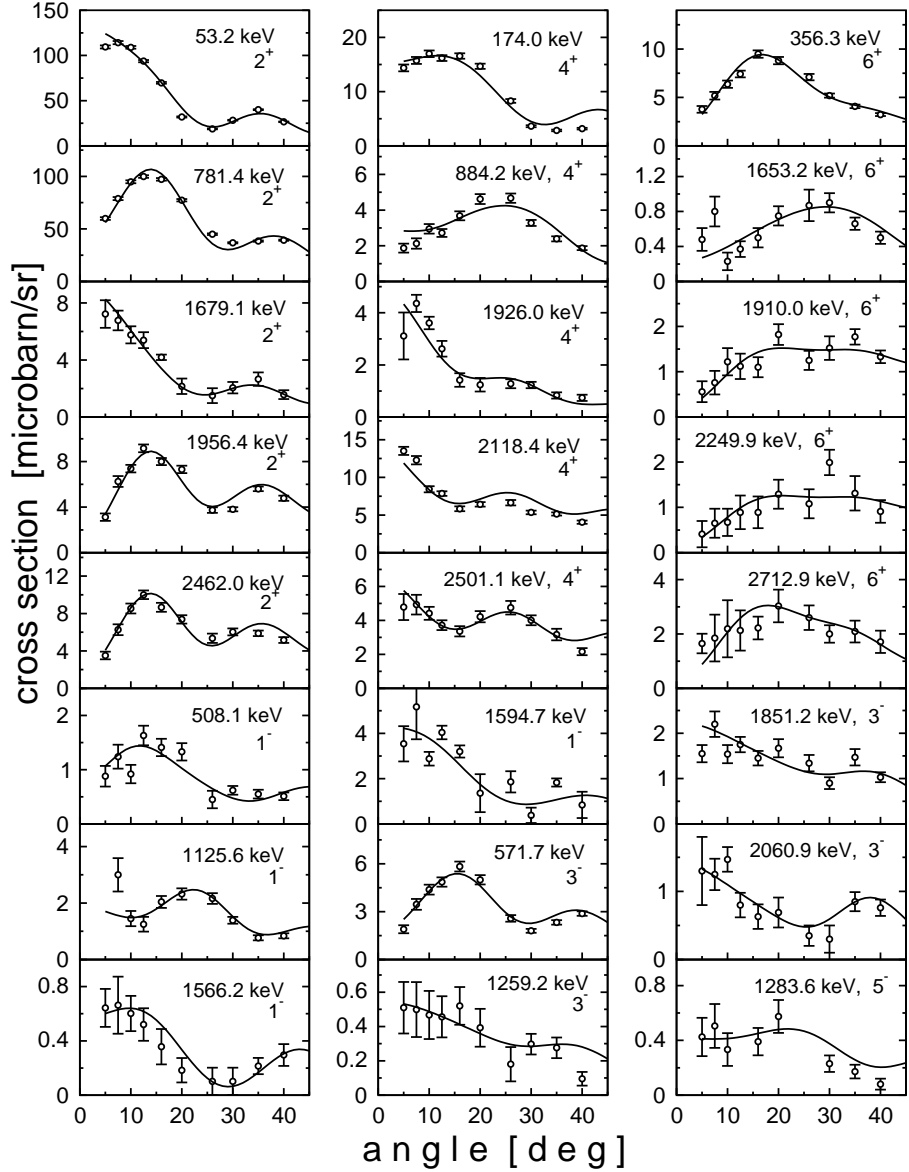}
    \caption{\label{fig:angl_distr_natur} Angular distributions of some
    excited states of natural parity and their fit by the CHUCK3
    calculations. The  $(ij)$ transfer configurations and schemes used
    in the calculations for the best fit are given in Table~\ref{tab:expEI}.}
\end{center}
\end{figure*}
%%%%%%%%%%%%%%%%%%%%%%%%%%%%%%%%%%%%%%%%%%%%%%%%%%%%%%%%%%%%%%%%%%%%%%

The angular distributions for the energies 952.6 and 1125.6 keV
are a special case. A state with spin $1^-$ is known at an energy
951.9 keV, however the angular distribution of the tritons at an
excitation energy of 952.6 keV is
 completely different from that for a known  $1^-$ state at 508.1 keV.
 Nevertheless, it can by fitted satisfactorily by an
 inclusion of  one-step and two-step excitations.
At the same time this angular distribution can also be fitted  by
a
 calculation for a $0^+$ excitation   mainly by
the $(1j_{15/2})^2$  transfer configuration  with a small
admixture of the $(1i_{11/2})^2$ transfer.  A spin of $0^+$ is
assigned
 to the state with closely lying energy of 927.3 keV in $^{232}$U,
 studied in the alpha
 decay of $^{236}$Pu \cite{Ard94}. Therefore,  we have not excluded spin
 assignment $0^+$,  though the present information is not sufficient to
 solve the problem. A similar angular distribution is observed for
 the excitation at 1125.6 keV.  But in this case the known level at the
 closely
lying energy of
 1127.8 keV has spin of $3^-$.   The CHUCK calculations give a
 satisfactory fit for  excitations of  the $1^-$ and $0^+$ states
 but not for $3^-$ state. It is interesting to note that the IBM
 calculations (see below) predict $0^+$ excitations at close energies
 for both states.

Thus we can make firm spin assignments for 16 states, relatively
firm assignments for 4 states and  tentative assignments for 4
states,
 in comparison with 14 states found in the preliminary analysis of the
 experimental data \cite{Wir04}.

It is difficult to estimate the additional number of $0^+$
excitations above 3 MeV from the spectra at 12.5$^\circ$ and
26$^\circ$ in Fig.~\ref{fig:spec_high}.  The intensities of the
lines corresponding to $0^+$ states have to be very low for
12.5$^\circ$ and much higher for 26$^\circ$. The lines fulfilling
this condition are labelled by their energies. Unfortunately, a
similar condition is fulfilled also for $6^+$ excitations as well
as for less probable $2^-$, $3^+$ states.  To be sure of the
presence of $0^+$ states in this high energy region at least a
spectrum for the most forward angle has to be measured. But there
is an impression from the measured spectra that the density of
$0^+$ states decreases for  energies above 3 MeV (or else that the
cross section of such excitations is very low and they are hidden
in very dense and complicated spectra). The maximum density of
$0^+$ states is observed in the interval between 2 and 3 MeV, as
one can see in Fig.~\ref{fig:strength}. The neutron pairing gap
for $^{230}$Th is about 750 keV  and  this leads to
two-quasiparticle excitations around 1.5 MeV. Thus the maximum
density  of $0^+$ states above this energy may be caused by
inclusion of such excitations, as predicted by the calculations
\cite{Rag76}.
%%%%%%%%%%%%%%%%%%%%%%%%%%%%%%%%%%%%%%%%%%%%%%%%%%%%%%%%%%%%%%%%%%%%
\begin{figure}
\begin{center}
\hspace{-14.0mm}
 \epsfig{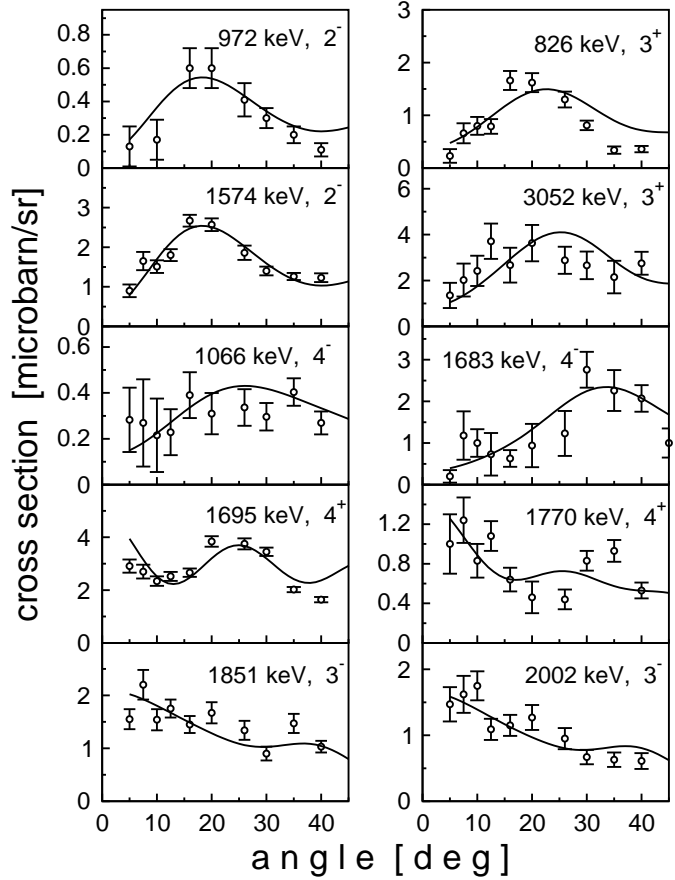}
    \caption{\label{fig:angl_distr_unnatur} Angular distributions and
    their fit by the CHUCK3 calculations for some
    excited states of unnatural parity, and for the states  where
    our assignments of spins are  in contradiction with the
    compilation \cite{Ako93}. The  $(ij)$ transfer configurations and schemes used
    in the calculations for the best fit are given in Table~\ref{tab:expEI}.}
\end{center}
\end{figure}
%%%%%%%%%%%%%%%%%%%%%%%%%%%%%%%%%%%%%%%%%%%%%%%%%%%%%%%%%%%%%%%%%%%%%%

Similar to  $0^+$ excitations,   the one-step transfer
calculations give a satisfactory fit of angular distributions for
about 70\% of the states with spins different from  $0^+$  but
about 30\% of these states need the inclusion of multi-step
excitations. Multi-step excitations have to be included to fit the
angular distributions already for the $2^+$, $4^+$ and $6^+$
states of the ground state band. Fig.~\ref{fig:schemes} shows the
schemes of the multi-step excitations tested for every state in
those cases where one-step transfer does not provide a successful
fit. Fig.~\ref{fig:angl_distr_natur} demonstrates the quality of
the fit of some  different-shaped angular distributions for the
excitation of states with  spins higher than $0^+$ by calculations
assuming one-step and one-step plus two-step excitations. The fits
for the ground state band are included in
Fig.~\ref{fig:angl_distr_natur}. Whereas natural parity states can
be populated by one-step or one-step plus two-step mechanisms, the
states of unnatural parity can be excited only by two-step
excitations. The angular distributions and their fits for such
excitations are presented in Fig.~\ref{fig:angl_distr_unnatur}.

The assignments of the spins resulting from such fits are
presented in Table~\ref{tab:expEI} together with other
experimental data. Detailed explanations of this table are given
in Sec. \ref{det_exp}.  Special comments are needed for the column
displaying the ratio $\sigma_{exp}/\sigma_{cal}$. Since we have no
a priori knowledge of microscopic structure of the excited states
and thus we do not know the relative contributions of the specific
$(j)^2$ transfer configurations to each of these states, these
ratios cannot be considered as spectroscopic factors.
Nevertheless, a very large ratio, such as in the case of the
$(1i_{11/2})^2$ transfer configurations used in the calculation
for the $0^+$ state at 635 keV, is unexpected.

Some additional comments on Table~\ref{tab:expEI} are needed.
There are some contradictions in the assignment of the energy of
the second $4^+$ level. It has been proposed from Coulomb
excitation to be at 772.1 keV in \cite{Ger84} and at 769.6 keV in
\cite{Kul89} and from the (d,pn$\gamma$) reaction at 775.5 keV
\cite{Ack94}. The value  769.6 keV is accepted in the compilation
\cite{Ako93}. We see a weak peak at 775.2 keV on the tail of the
very strong  peak at 781.4 keV as confirmation of the 775.5 keV
assignment \cite{Ack94}. There is not even a hint of a peak at
769.6 keV. Again the $8^+$ and $10^+$ levels proposed in
\cite{Kul89} at 1251.4 keV and 1520.4 keV (both accepted in the
compilation \cite{Ako93}) are in contradiction with the assignment
of the $8^+$ level at 1243.2 keV and with the calculated energy of
the $10^+$ level at 1487 keV \cite{Ack94}.  The (p,t) study
confirms the assignment for the $8^+$ level at 1243.2 keV by the
observation of a weak peak at 1241.2 keV and no peak in the
vicinity of 1251.4 keV. After this confirmation the smooth change
of the inertial parameter in the band  prefers the energy 1487 keV
for the $10^+$ level and rejects the energy of 1520.4 keV.

There are several levels in $^{230}$Th  for which spins $1^{(-)}$,
$2^+$ or $1,2^+$  were assigned   from the $\beta^-$-decay and/or
inelastic scattering and for which the (p,t) angular distributions
are measured: 1573.5, 1695.7, 1770.7, 1839.6, 1849.6, 1966.9,
1973.4, 2000.9,  2010.1 and 2024.7 keV (energies are from the
compilation \cite{Ako93}).  Angular distributions from the (p,t)
reaction confirm spin $2^+$ for some of them: 1840.0, 1967.1,
1972.0, 2010.3 and 2025.6 keV (energies as determined from the
(p,t) reaction). The (p,t) angular distributions for other states
cannot be fitted for the spins given in \cite{Ako93}. If the
energies 1574.5(6) keV from the (p,t) reaction and 1573.5(2) keV
from \cite{Ako93} correspond to the same level,  then a different
assignment is suggested  by the (p,t) angular distribution to this
state as $2^-$ or $3^+$. In any case a $2^+$ excitation would
manifest itself in the (p,t) angular distribution and therefore
has to be excluded. For the states at 1694.9 and 1769.6 keV we
observe angular distributions which can be fitted by calculations
for $4^+$ states (or even for a $0^+$ state plus a constant of
about 2~$\mu$b for the state of 1694.9 keV). Either way, the
assignment $2^+$ can be excluded. Finally, the (p,t) angular
distributions for the states at 1851.4 and 2001.6 keV prefer an
assignment of $3^-$ in contradiction to $(2^+)$ and $1,2^+$,
respectively, as given in \cite{Ako93}.

%%%%%%%%%%%%%%%%%%%%

\begin{figure}
\begin{center}
\epsfig{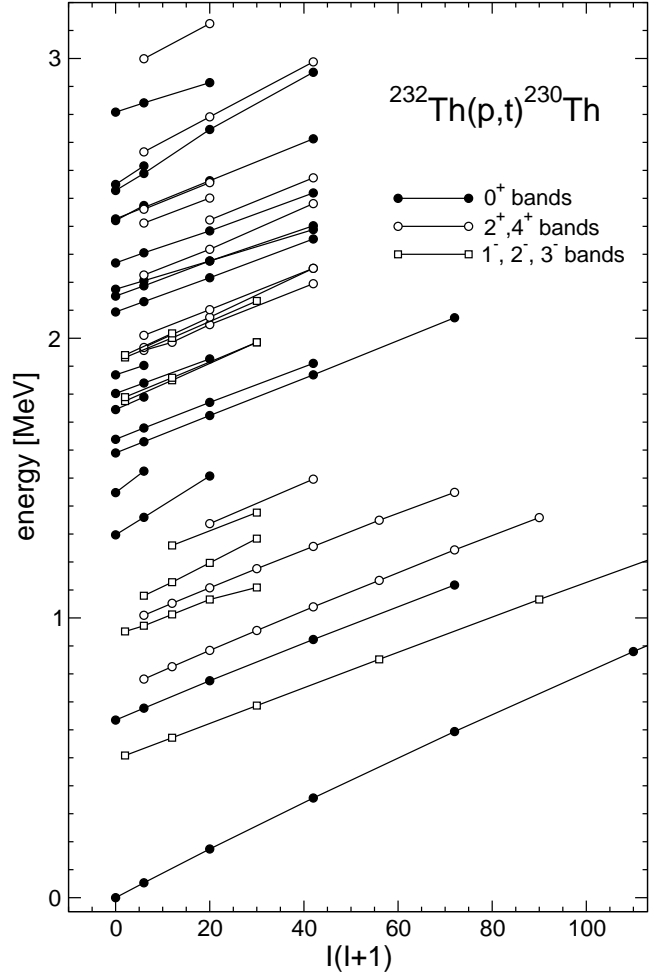}
\caption{\label{fig:bands} Collective bands based on the $0^{+}$,
$2^{+}$, $4^{+}$, $1^{-}$,  $2^{-}$ and $3^{-}$ excited states in
$^{230}$Th as assigned from the DWBA fit of the angular
distributions from the (p,t) reaction.}
\end{center}
\end{figure}

%%%%%%%%%%%%%%%%%%%%%%%%%%%%%%%%%%%%%%%%%%%%%%%%%%%%%%%%%%%%%%%%%%%%%%%

\newcolumntype{d}{D{.}{.}{3}}
\begin{table*}[]
\caption{\label{tab:bands} \normalsize{The sequences of  states
qualifying as candidates for rotational bands (from the CHUCK fit,
the (p,t) cross sections and the inertial parameters). More
accurate values of energies are taken from the first two columns
of Table~\ref{tab:expEI}. The energies taken in brackets
correspond to the sequences  assigned tentatively. }}
\begin{ruledtabular}
\begin{tabular}{cccccccccc}
\smallskip\\
 $0^+$ & $1^+$ & $2^+$ & $3^+$ & $4^+$ & $5^+$ & $6^+$ & $7^+$
& $ 8^+$ &
\smallskip\\
\hline
\smallskip\\
0.0 &&    53.2 &&     174.0 &&   356.3 &&   593.8 &\\
634.9 && 677.5 && 775.5 && 923.0 && 1117.5 &\\
    &&   781.4 & 825.6 & 883.9 & 955.1 & 1039.6 & 1134.2 & 1243.3 & \\
    &&   1009.6 & 1052.3 & 1107.5 & 1176.1 & 1255.5 & 1349.3 & 1448.7 &\\
1297.1&& 1359.5 && 1507.4 &&&&&\\
       &&&& 1337.2 && 1496.0 &&  &\\
1447.9 && 1524.8 &&&&&&&\\
1589.8 && 1630.1 && 1723.5 && 1868.9 && 2073.2 &\\
1638.5 && 1679.1 && 1770.7 && 1910.0 &&&\\
(1744.9) && 1789.4 &&&&&&&\\
1802.5 && 1839.6 && 1926.0 &&&&&\\
(1868.9) && 1902.7 &&&&&&&\\
     && 1956.4 & 1985.4 & 2048.7 && 2194.8 &&&\\
     && 1966.9 && 2074.9 && 2249.9 &&&\\
     && 2010.1 && 2102.0 && 2249.9 &&&\\
2093.9 && 2130.7 && 2216.0 && 2354.8 &&&\\
2150.5 && 2187.1 && 2276.0 && 2402.0 &&&\\
2175.1 && 2205.4 && 2276.0 && 2388.4 &&&\\
       && 2226.0 && 2317.7 && 2481.3 &&&\\
2268.9 && 2305.4 && 2383.8 && 2519.3 &&&\\
(2422.7) && 2474.3 &&&&&&  &\\
       &&&& 2422.7 && 2573.2 &&  &\\
(2426.4) && 2467.2 && 2562.9 && 2712.9 &&&\\
       && 2411.6 && 2501.1 &&&&&\\
       && 2461.0 && 2556.2 &&&&&\\
2528.1 && 2589.1 && 2746.2 && 2950.5 &&&\\
2549.8 && 2616.0 &&&&&&&\\
       && (2666.0) && 2791.5 && 2987.9 &&&\\
2808.1 && 2841.3 && 2913.6 &&&&&\\
       && 2999.0 && 3124.7 &&&&&\\
\hline
\smallskip\\
& $1^-$ & $2^-$ & $3^-$ & $4^-$ & $5^-$ & $6^-$ & $7^-$ & $8^-$
&$^9-$
\smallskip\\
\hline
\smallskip\\
& 508.2 && 571.7 && 686.7 && 851.9 && 1065.9\\
& 951.9 & 972.1 & 1012.5 & 1065.9 & 1109.0 &&&&\\
       && 1079.4 & 1127.8 & 1196.8 & 1283.6 &&&&\\
&      && (1259.2) && 1376.7 &&&&\\
& (1789.4) && 1858.6 && 1985.4 &&&&\\
& (1775.2) && 1849.6 && 1985.4 &&&&\\
& (1931.1) && 2000.9 && 2133.2 &&&&\\
& (1939.8) && 2017.3 &&&&&&\\
\end{tabular}
\end{ruledtabular}
\end{table*}

%%%%%%%%%%%%%%%%%%%%%%%%%%%%%%%%%%%%%%%%%%%%%%%%%%%%%%%%%%%%%%%%%%%%%

\subsection{\label{sec:bands}Collective bands in $^{230}$Th}

After the assignment of spins to all excited states the sequences
of  states can be distinguished which show the characteristics of
a rotational band structure. An identification of the states
attributed to rotational bands was made on the following
conditions:

a) the angular distribution for a state as  band member candidate
is fitted by the DWBA calculations for the spin necessary to put
this state in the band;

b) the transfer cross section in the (p,t) reaction  to the states
in the potential band has to decrease with increasing  spin;

c) the energies of the states in the band can be fitted
approximately by the expression for a rotational band   $E = E_0+
AI(I+1)$ with a small and smooth variation of the inertial
parameter $A$.
 Collective bands identified in such a way are shown
in Fig.~\ref{fig:bands} and are listed in Table~\ref{tab:bands}
(for a calculation of the moments of inertia). The procedure can
be justified in that some sequences meeting the above criteria are
already known from gamma-ray spectroscopy to be rotational bands,
so other similar sequences are very probably rotational bands too.
The straight lines in Fig.~9 strengthen the argument for these
assignments. For example the mean deviation of the experimental
energies from the calculated rotational values for the longest
newly assigned band based on the state 0$^+$ at 1589 keV is only
1.0 keV; for the band based on the 0$^+$ state at 2093.9 keV it is
1.3 keV; and for the band above the 0$^+$ state at 2268.9 keV it
is 3.9 keV. Even for the band above the 0$^+$ state at 2426.4 keV
assigned tentatively the deviation is less than 1 keV. The
observed deviations are all consistent with the stretching effect
typical for rotational bands. Nevertheless additional information
(on E2 transitions at least) is needed to confirm these
assignments.

It is worth mentioning that the tentative assignment $0^+$ for the
state at 2426.4 keV is supported by a sequence of 3 other states
($2^+$, $4^+$, $6^+$) on top of it. Three other tentative $0^+$
states (1745.3, 1868.9, and 2422.7 keV) have only one tentative
state ($2^+$) on top of them and the band sequence is not based on
$\gamma$ - ray transition but on energy arguments. In Table
\ref{tab:moments} we present moments of inertia (MoI) obtained by
fitting the level energies of the bands displayed in
Fig.~\ref{fig:bands}  by the expression $E = E_0 + AI(I+1)$. In
upper part of the table  those sequences are presented which are
connected by known $\gamma$ - ray transitions or have at least 3
levels and in lower part of the table the sequences having only 2
levels or tentatively assigned are presented. The obtained MoI
cover a broad range, from ~$\sim$50 MeV$^{-1}$ to ~$\sim$100
MeV$^{-1}$. The negative parity bands based on the states with
spin 1$^-$ interpreted as the octupole-vibrational bands
\cite{Ack94} have high MoI (the 1$^-$ band at 951 keV has the
largest). The $0^+$ band at 1297 keV, considered as $\beta$ -
vibrational band, has the smallest MoI. At this stage, it is
difficult to make a complete correlation between the intrinsic
structure of the bands and the magnitude of their MoI.
Nevertheless one can assume also for the $0^+$ bands that the
largest MoI could be related  to the octupole phonon structure and
the smallest MoI could be related to the one-phonon quadrupole
structure. The bands with the intermediate values of the MoI could
be based on the two-phonon quadrupole excitations.

If the moments of inertia do indeed carry information on the inner
structure of the bands, then the numbers of excitations with
different structure are comparable. This would be in contradiction
with the IBM calculation which predicts predominantly the octupole
two-phonon structure of 0$^+$ excitations (see below). The nature
of 0$^+$ excitations, derived from calculations in the framework
of the QPM as predominately quadrupole, is in contradiction with
both the IBM calculation and with the above mentioned empirical
observation (see below).

%%%%%%%%%%%%%%%%%%%%%%%%%%%%%%%%%%%%%%%%%%%%%%%%%%%%%%%%%%%%%%%%%%%%
\newcolumntype{d}{D{.}{.}{2}}
\begin{table}
\caption
    {\label{tab:moments} Moments of inertia for the bands in $^{230}$Th
as assigned from the angular distributions
  from the $^{232}$Th(p,t)$^{230}$Th reaction. Results derived from
  the sequences having only 2 levels or assigned tentatively are
  given in lower part of the table.}
 \begin{ruledtabular}
    \begin{tabular}{cccccc}
    \smallskip\\
E [keV] & J(0$^+$) & E [keV] & J(2$^+$) & E [keV] & J($1^-,3^-$)\\
\smallskip\\
\hline
\smallskip\\
0.0 & 56.8 & 781  & 67.5 & 508 & 78.3\\
635 & 70.1 & 1009 & 70.0 & 951 & 98.5\\
1297& 47.8 & 1956 & 62.9 & 1079 & 61.7\\
1589& 74.0 & 1967 & 64.5 & 1259 & 79.4\\
1639& 75.7 & 2010 & 75.7 & 1789 & 72.0\\
1802& 80.5 & 2226 & 76.0 & 1931 & 71.3\\
2093& 81.2 & 2666 & 55.5 &&\\
2150& 81.6 &&&\\
2175& 98.5 &&&\\
2269& 81.8 &&&\\
2426& 74.0 &&&\\
2528& 49.0 &&&\\
2808& 89.5 &&&\\
&&&&\\
1448& 40.0 & 2412 & 78.0 & 1259 & 79.4\\
1745& 67.2 & 2461 & 73.1 &&\\
1868& 88.4 &  2999 & 55.4 &&\\
2422& 58.0 &&&\\
2550& 45.7 &&&\\
   \end{tabular}
\end{ruledtabular}
\end{table}
%%%%%%%%%%%%%%%%%%%%%%%%%%%%%%%%%%%%%%%%%%%%%

%%%%%%%%%%%%%%%%%%%%%%%%%%%%%%%%%%%%%%%%%%%%%%%%%%%%%%%%%%%%%%%%%%%%%%

\subsection{Excited states with spins higher than $0^+$}

Other states, mainly with spins $2^+$, $4^+$ and $6^+$ are
intensively excited in  the (p,t) reaction. The nature of these
states may only be assumed. Some of these states could belong to
the collective bands based on $0^+$ states. Some of the  $2^+$
states could be quadrupolar (one-phonon) vibrational states with
correspondence to $0^+$ excited states, since in deformed nuclei
every excitation of  angular momentum $I^{\pi}$ splits into states
distinguished by their $K$ quantum numbers ranging from $0$  to
$I$. Some $4^+$ excited states could be hexadecapole vibrational
excitations and  $2^+$ and $0^+$ states should correspond to this
class of states for the same reason. If this speculation reflects
reality then the number of $0^+$ states has to be the largest.
However, the observation is in contradiction with this conjecture
unless very weak $0^+$ excitations are not seen in the (p,t)
reaction. Attributing the underlying structure to each of the
observed states is not possible with the presently available
experimental data. To give at least a hint for the structure of
these states the experimental energy distribution of the (p,t)
transfer strength for the 0$^+$, 2$^+$ and 4$^+$ excitations is
plotted in Fig.~\ref{fig:strength}.

\begin{figure}
\begin{center}
\epsfig{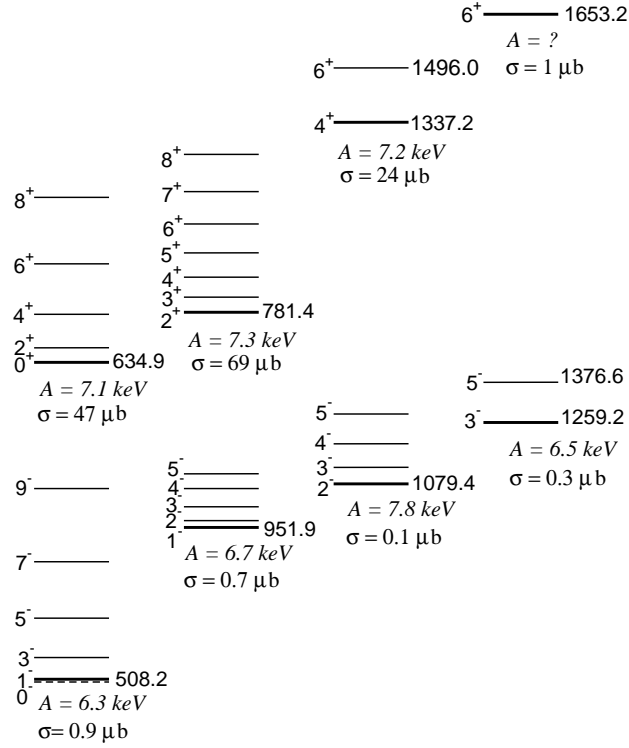}
\caption{\label{fig:quadruplet} Assumed multiplets of states of
the octupole one-phonon  (bottom) and  the octupole two-phonon
(top) excitations  and corresponding collective bands.}
\end{center}
\end{figure}

A most remarkable feature in Fig.~\ref{fig:strength} is the most
strongly excited 4$^+$ state at 1337 keV, which can be related to
the strongly excited 0$^+$ state at 635 keV and 2$^+$ state at 781
keV. The inertial parameters derived from the bands based on these
states are practically the same. The only explanation is that
these states have the same structure. One can assume that these
states are a triplet originating from an excitation of
multipolarity 4 as in the case of a quadrupole two-phonon
excitation. However, a corresponding strongly excited one phonon
quadrupole excitation with spins 0$^+$ and 2$^+$ are not observed
at lower energy. Indeed, the 0$^+$ state at 1297 keV, which was
identified as the $\beta$-vibrational state \cite{Ack94}, is
rather weak. Thus quadrupole two-phonon excitations have to be
excluded. On the assumption that these states have a two-phonon
octupole structure a quadruplet of states has to be observed with
spins from 0$^+$ to 6$^+$. Experimentally a state with spin 6$^+$
is identified at 1653 keV, although its excitation is only of 0.9
$\mu$b, i.e. much weaker than the other states. If this assumption
is correct, then the interpretation of the first  0$^+$ state as a
two-phonon excitation obtains some confirmation. This quadruplet
of states and the corresponding bands are displayed in
Fig.~\ref{fig:quadruplet}. Note however, that the moments of
inertia derived from the band based on the states of the multiplet
are somewhat smaller than the ones derived for the bands based on
negative parity states (Table~\ref{tab:moments}).

An indirect confirmation of the above assumption would be the
existence of another quadruplet. The $1^-$ levels at 508 keV and
952 keV and the $2^-$ level at 1079 keV were interpreted as
members of a quadruplet of octupole shape oscillations with $K^\pi
= 0^-$ to $3^-$  \cite{Ack94}. The  $3^-$ level at 1259 keV
identified from the (p,t) reaction in the present study can be the
missing member of this quadruplet, as shown in
Fig.~\ref{fig:quadruplet}. The energy separation between the
$K^\pi = 0^-$ and $K^\pi = 1^-$ and between the $K^\pi = 1^-$ and
$K^\pi = 2^-$ band heads differ strongly. This was explained by a
strong coupling of the bands based on the last two states
\cite{Ack94}. The energy separation between the  3$^-$ level at
1259 keV and the $2^-$ level is close to the one between the
second $1^-$ and $2^-$ levels, which can also be attributed to the
coupling of these three states.
%%%%%%%%%%%%%%%%%%%%%%%%%%%%%%%%%%%%%%%%%%%%%%%%%%%%%%%%%%
%%%%%%%%%%%%%%%%%%%%%%%%%%%%%%%%%%%%%%%%%%%%%%%%%%%%

\section{\label{Disc} Discussion}

\subsection{$0^+$ excitations}

The importance of pairing in the enhancement of the cross section
of the $0^+$ two-nucleon transfer reaction  was noted already in
an earlier publication \cite{Bes66,Bro73}. The superfluid  ground
states of deformed nuclei are strongly populated in the (p,t)
reaction due to the large overlap of the wave functions between
nuclei with neutron number $N$ and $N\pm2$. The excited $0^+$
states can be populated  in the (p,t) reaction due to the
fluctuation of the pairing field. Such states represent pairing
vibration modes \cite{Bes66,Bro73}. It was realized that pairing
correlations in deformed nuclei are induced not only by monopole
but  also by quadrupole pairing interactions. If  both pairing
fields are comparable in strength and if there is a nonuniform
distribution of oblate and prolate single-particle orbitals around
the Fermi surface, this may give rise to $0^+$ excitations treated
as pairing isomers decoupled from the deformed superfluid ground
state \cite{Bro72,Rag76}.  An asymmetry between (p,t) and (t,p)
cross sections predicted by this model was confirmed in an
experiment by Casten {\it et al.} \cite{Cas72}.

The structure of excited $0^+$ states in deformed even-even nuclei
is still a matter of controversial discussion despite intensive
investigation. Traditionally the first excited $0_2^+$ state has
been interpreted as the beta-vibrational excitation of the ground
state. However, in many nuclei the $0_2^+$ state has only weak
transitions to the ground-state band, while strong electric
quadrupole transitions to the gamma band have been found
\cite{Cas94}. This contradicts the traditional interpretation,
since a transition from a beta-vibrational state to the gamma band
is suppressed due to the destruction of a beta phonon, and, at the
same time, the creation of a gamma phonon. The unclear situation
led to an intense debate about the structure of low-lying $0^+$
states.

Maher {\it et al.} \cite{Mah72} were the first who noticed an
interesting feature of $0^+$ excited states in the actinides. The
strong excitations of the first excited $0^+$ states in the (p,t)
reaction, combined with all other available evidence (rather weak
E2 transitions to the ground-state band, strong $\alpha$ decays
leading to them, the strong  Coulomb excitation of the associated
collective bands) suggest that these states represent a new and
stable collective excitation, different in character from the most
common formulation of the pairing vibration as well as from the
$\beta$ vibration usually found in the deformed rare-earth nuclei.
The second excited $0^+$ states in actinides (firmly assigned)
demonstrate completely different features
\cite{Ack94,Bal95,Bal96}. Weak excitation in the (p,t) reaction,
relatively strong $E2$ transitions to the ground-state band and a
small $B(E1)/B(E2)$ ratio for transitions to $2^+$ and $1^-$
states give evidence that  they could be the usual $\beta$
vibrational states. For $^{230}$Th this is the case for the level
at 1297.8 keV \cite{Ack94}. Otsuka and Sugita \cite{Ots89} applied
the $spdf$-interacting boson model  to the actinide nuclei, aiming
to get a unified description of quadrupole-octupole collective
states. They suggested the first excited $0^+$ band be referred to
as the "super $\beta$ band" thus emphasizing the difference of the
structure of this state from that of the usual beta-vibrational
state. They also predicted the existence of the second excited
$0^+$ band to be the usual $\beta$ band that was confirmed later.
There are evidences for the first 0+ state in $^{230}$Th
(quadruplet of states, large moment of inertia comparable to that
for the octupole-vibrational bands, together with the features
noticed by Maher {\it et al.} \cite{Mah72} and listed above) to
carry the two-phonon octupole nature. However, the B(E1)/B(E2)
ratio for transitions to the 1- and 2+ states is even smaller for
this state than for the state at 1297 keV: $\sim 4 \cdot 10^{-7}$
fm$^{-2}$ compared to $\sim 7 \cdot 10^{-7}$ fm$^{-2}$
\cite{Ack94}. Intuitively one would expect that a large
B(E1)/B(E2) ratio might be characteristic for a
two-octupole-phonon excitation, whereas  a small ratio might
indicate a  $\beta$ shape oscillation. That is true for the state
at 1297 keV (which indeed is a beta vibrational state), but not
true for the first excited state. Moreover, the IBM and the QPM
predict one phonon quadrupole nature for this state (see below).
Therefore the available data do not allow the firm conclusion on
the nature of this state.

Understanding of the structure of the higher excited states
remains a challenge for further  experimental studies (e.g.
$\gamma$ spectroscopy in the (p,t$\gamma$)-reaction) and for
nuclear theory. The first attempt to explain experimental data of
a large number of $0^+$ excited  states in $^{158}$Gd \cite{Les02}
was a phenomenological approach \cite{Zam02} based on the extended
interacting boson model ({\it spdf}-IBM), which accounted for a
large fraction of the observed states. The importance of the
octupole degrees of freedom was revealed. The first microscopic
approach was performed in the framework of the projected shell
model (PSM) \cite{Sun03}, using a restricted space spanned by two
and four quasiparticle states. The IBM calculation reproduced
satisfactorily all energy levels in $^{158}$Gd and gave small $E2$
decay strengths for them. Soloviev and co-workers \cite{Sol89,
Sol949697} applied the quasiparticle-phonon model (QPM) to get a
microscopic understanding of low-lying $0^+$ states. The QPM was
also applied to $^{158}$Gd \cite{Lo04}.  It predicts an sizable
fraction of the $0^+$ states to have large or dominant two-phonon
components, mainly built from collective octupole phonon
components, in agreement with the IBM calculations \cite{Zam02}.

The results of a specific analysis of experimental data for
actinide nuclei were compared to  the {\it spdf}-IBM calculations
in our previous paper \cite{Wir04}. After publication of these
data, Lo Iudice {\it et al.} carried out a calculation for these
nuclei in the framework of  the QPM \cite{Lo05}.   Although
energies, $E2$ and $E0$ transition strengths and two-nucleon
transfer spectroscopic factors were computed,  only some of them
could be compared with the then available experimental data. We
compare our new data for $^{230}$Th with the results of both
calculations below.

No extended calculations have yet been carried out  for the
excitation of the two-quasiparticle (2QP) modes, which are
expected to occur at excitation energies of about twice the
pairing gap energy, i.e. about 1.5 MeV in our case. Only very
restricted microscopic calculations for low energies were
attempted by Ragnarsson and Broglia \cite{Rag76}. It would be
desirable to extend these calculations to higher energies. Just
above 1.5 MeV, exactly in the range 2.0--2.5 MeV, a bump in the
distribution of the (p,t) transfer strength is observed
(Fig.~\ref{fig:strength}). At least some of these excitations
could be of 2QP-nature. A model by Rij and Kahana \cite{Rij71}
describing the $0^+$ state as a pair of holes in the oblate
1/2[501] Nilsson level should also be mentioned.

The monopole pairing vibration state for neutron-pair excitation
(n-MPV) is expected to be strongly excited  in the (p,t) reaction
because of the large overlap of the wave function of such a state
with that of the target nucleus  ground state. A dominant  (p,t)
cross section for a single state in the spectrum  (besides the
ground and the first excited states) is observed in some nuclei.
In $^{229}$Pa the cross section for the $L=0$ transfer to the
state at 1500  keV is about 15\,\% of that for the ground state
and comparable to that for the first excited $0^+$ state in
$^{230}$Th \cite{Lev94}. However, this behavior  is unexpected,
since this energy corresponds more to that of the proton-pair
excitation (p-MPV), where the cross section  is expected to be
much weaker. At the same time, a dominant cross section in
$^{228}$Th is observed for a single state at about 2.1 MeV,  much
higher than in $^{229}$Pa, which can be considered as being due to
the $^{228}$Th core plus a proton. These facts, as well as
practically no correspondence of the energy distribution of the
(p,t) strength in these two nuclei, need a theoretical
explanation. No dominant excitation at higher energies is observed
in $^{230}$Th. In the case of a relatively dense spectrum of $0^+$
states, fragmentation of the n-MPV state to nearby states is
possible. Such a group of states in $^{230}$Th  around 1600 keV
could be a result of such fragmentation. The summed cross section
of this group is about 13\,\% of that for the ground state. A
similarly dense spectrum occurs also in $^{228}$Th and $^{229}$Pa;
nevertheless,  these nuclei demonstrate dominant excitation of
individual states. Unfortunately, no calculations have been
carried out for odd nuclei, though they are planned for $^{229}$Pa
\cite{Sushk}. There is no clear picture concerning the problem of
the MPV as well as of the 2QP states.

 %%%%%%%%%%%%%%%%%%%%%%%%%%%%%%%%%%%%%%%%%%%%%%%%%%%%%%%%%%%%%%%%%
\begin{figure}[h]
\begin{center}
\epsfig{file=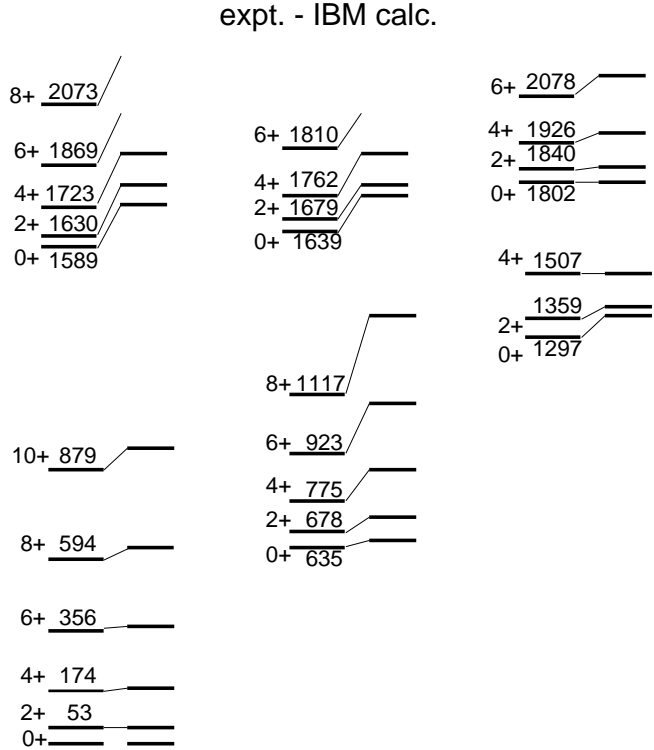, width=8.6cm,angle=0}
    \caption{\label{fig:exp_IBM} Lowest $0^+$ bands identified earlier and
    suggested in this analysis  in comparison to
    the {\it spdf}-IBM calculations.}
\end{center}
\end{figure}
%%%%%%%%%%%%%%%%%%%%%%%%%%%%%%%%%%%%%%%%%%%%%%%%%%%%%%%%%%%%%%%%
%%%%%%%%%%%%%%%%%%%%%%%%%%%%%%%%%%%%%%%%%%%%%%%%%%%%%%%%%%%%%%%%
\begin{figure*}[!]
\begin{center}
\epsfig{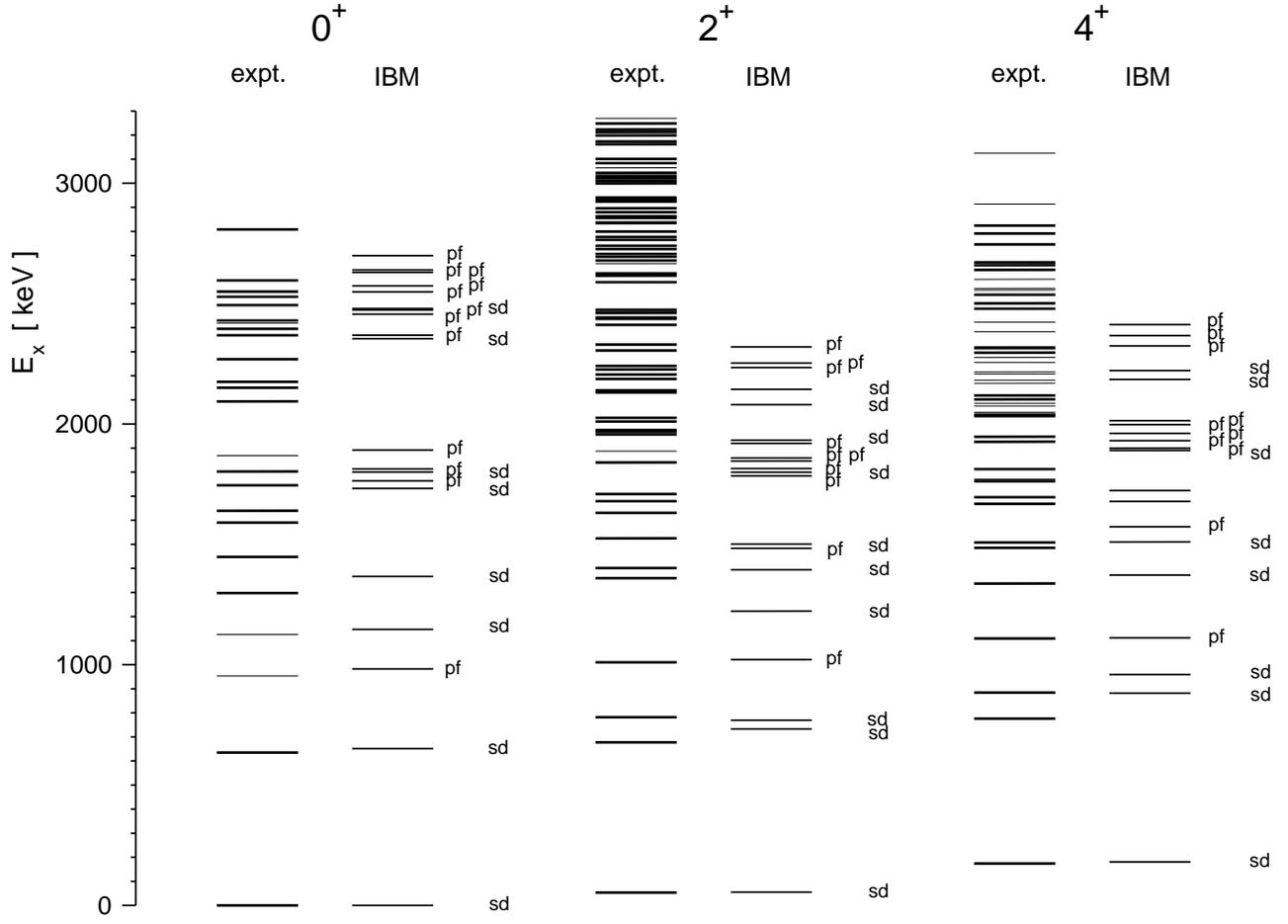}
    \caption{\label{fig:compil_zeroIBM} Energies of all experimentally
    assigned excited $0^+$, $2^+$ and $4^+$ states in $^{230}$Th
    in comparison with {\it spdf}-IBM calculations. The structure of the states
    derived
    from calculations is indicated to the right of the corresponding lines.
    The firmly assigned states are shown by thick lines, while  tentatively
    assigned states are indicated by thin lines.}
\end{center}
\end{figure*}
%%%%%%%%%%%%%%%%%%%%%%%%%%%%%%%%%%%%%%%%%%%%%%%%%%%%%%%%%%%%%%%%%%

    \subsection{IBM calculations}

Although $^{230}$Th  is considered a vibrational-like nucleus, the
inclusion of  the octupole degree of freedom in the description of
its properties turned out to be important \cite{Ack94}. The role
of the octupole degree of freedom in deformed actinide nuclei and
the related description with  $f$ bosons added to the IBM in the
$sd$ boson space ($sdf$-IBM) has
 been studied in \cite{Cot98}. Despite reproducing  reasonably well the
 main features of the observed low-lying negative parity states in the
 rare earth nuclei \cite{Cot96},  the $sdf$-IBM was not so successful for the
 actinide nuclei. A better reproduction of the relevant data is
 obtained if a $p$ boson is included in addition to the $f$ boson
 without seeking an understanding of its physical nature \cite{Zam0103}.

The IBM Hamiltonian in the $spdf$ space includes a vibrational
contribution and a quadrupole interaction in the simple form
\cite{Zam02,Zam0103}.
 \begin{equation}\label{eq:IBM-Ham}
H = {\epsilon}_d {\hat{n}_d} + {\epsilon}_p {\hat{n}_p}
  + {\epsilon}_f {\hat{n}_f} - {\kappa} {\hat{Q}}_{spdf} \cdot
{\hat{Q}}_{spdf},
\end{equation}
where ${\epsilon}_d$ , ${\epsilon}_p$, and
 ${\epsilon}_f $ are the boson energies and $\hat{n}_d$, $\hat{n}_p$,
and $\hat{n}_f $ are the boson number operators. Note that the
same strength  ${\kappa}$ of the quadrupole interaction describes
the $sd$ bosons and the $pf$ bosons. The $\hat{Q}_{spdf}$
quadrupole operator has the form
\begin{eqnarray}\label{eq:Qoperator}
\lefteqn{\hat{Q}_{spdf} = \hat{Q}_{sd}+  \hat{Q}_{pf} =
[s^{\dag}\tilde{d}  + d^{\dag}{s}]^{(2)}}\nonumber\\
&&{}-  {\chi_{sd}}[d^{\dag}\tilde{d}]^{(2)}
+ (3/5) \sqrt{7} [p^{\dag}\tilde{f}  + f^{\dag}\tilde{p}]^{(2)}\nonumber\\
&&{}- (9/10) \sqrt{3} [p^{\dag}\tilde{p}]^{(2)} - (3/10) \sqrt{42}
[f^{\dag}\tilde{f}]^{(2)}
\end{eqnarray}
The  factor in front of the $[d^{\dag}\tilde{d}]^{(2)}$ term may
be adjusted by introducing an additional parameter ${\chi}_{sd}$.

The IBM parameters in the $sd$ boson space are determined  by the
low energy spacing of the ground state band  and the $I^{\pi} =
2^+_1 , K^{\pi} = 0^+$  and $I^{\pi} = 2^+_1, K^{\pi} = 2^+ $ band
heads, respectively. The $pf$ boson parameters are chosen to
reproduce the $K^{\pi} = 0^-$ and  $K^{\pi} = 1^- $ band heads;
they are determined by the experimental energies of the $I^{\pi} =
1_1^-, K^{\pi} = 0^-$ and $I^{\pi} = 1_2^-, K^{\pi} = 1^-$ states.
The total number of bosons is 11. The number of negative parity
bosons is allowed to range from 0 to 3. Relative to the
calculations presented in our earlier paper \cite{Wir04}, only
some parameters are slightly changed in the present calculations:
${\epsilon}_p$=-1.00, ${\epsilon}_d$=0.25, ${\epsilon}_f$=-0.90,
${\kappa}$=-0.014, ${\chi}_{sd}$=-1.00. As for rare earth nuclei
\cite{Zam02}, in the $spdf$-IBA calculations mixing between $d$
and $pf$ bosons is neglected and the $f$ (and $p$) bosons account
for octupole collectivity. In Fig.~\ref{fig:exp_IBM}  we display
the firmly assigned experimental $0^+$ states  and the sequences
on top of them, in comparison with the  $spdf$-IBM calculations.

The full experimental spectrum of the $0^{+}$ states in
$^{230}$Th, including relatively firm and tentative assignments,
and the results of $spdf$-IBA calculations are compared in
Fig.~\ref{fig:compil_zeroIBM}. In  the energy ranges covered
experimentally, the IBM  predicts  seven  excited $0^{+}$ states
of pure $sd$ (quadrupolar) bosonic structure, and twelve excited
$0^{+}$ states which have two bosons in the  $pf$ boson space.
They could be related to octupole two phonon excitations (OTP).
The nature of the states according to the calculations is
indicated in Fig.~\ref{fig:compil_zeroIBM}. The number and the
nature of the 0$^+$ states in the IBM calculations depends on the
parameter values but the solution presented in
Fig.~\ref{fig:compil_zeroIBM} is stable for a realistic variation
of these parameters. A complete study of this dependencies would
be instructive but it is not the goal of the present paper.

We can see a reasonable correlation in excitation energy between
experiment and calculation up to 1.4 MeV. The calculation predicts
the first excited 0+ state as a quadrupole excitation in the sd
space at an energy close to the experimental state at 635 keV. The
second excited 0+ state is predicted as an octupole two-phonon
excitation and could correspond to the tentative assigned 0+ state
at 953 keV. The following two are predicted to have a quadrupole
structure and correspond energetically to the tentative assigned
0+ state at 1126 keV and to the 0+ state at 1297 keV,
respectively. The latter was, indeed,  considered the $\beta$
-vibrational state in Ref. \cite{Ack94}.

Thus the IBM calculation with the parameterizations used predicts
for $^{230}$Th 20 excited  $0^{+}$ states in the  energy range
below 2.7 MeV. Accounting in addition  for the presence of
monopole pairing vibrational states, two-quasiparticle states and
perhaps a state from hexadecapole collectivity, not included in
the calculation, we can consider  24 observed $0^+$ excitations as
nearly perfect agreement with calculated  number of such
excitations. But there is no clarity concerning the nature of
these excitations without additional experimental information. The
IBM also fails completely  to reproduce the (p,t) spectroscopic
factors. The calculated first excited $0^{+}$ state occurs with
about one percent of the transfer strength of the ground state,
and the higher states are even  weaker, whereas experimentally the
excited states show at least about 70\,\% of the ground state
transfer strength.

A calculation in the framework of the {\it spdf}-IBM  gives 20
levels of spins 2$^+$, 4$^+$ and 6$^+$ in comparison with 40, 32
and 11 identified levels of these spins in the energy region below
2.7, 2.3 and 2.4 MeV, respectively. The spectrum of $2^+$ and
$4^+$ states in comparison with the {\it spdf}-IBM calculation is
given in Fig.~\ref{fig:compil_zeroIBM}. Most of the 6$^+$ states
could not be observed in the (p,t) reaction because of their low
cross section. As far as 2$^+$ and 4$^+$ states are concerned, the
number of experimental levels is much higher than the prediction
of the {\it spdf} IBM model.

\subsection{QPM calculations}

 The ability of the QPM to describe multiple
$0^+$ states (energies, $E2$ and $E0$ strengths, two-nucleon
spectroscopic factors) was demonstrated for $^{158}$Gd
\cite{Lo04}. An extension of the QPM to describe the $0^+$ states
in the actinides \cite{Lo05} was made  after our publication on
the results of  a preliminary analysis of the experimental data
\cite{Wir04}. The present calculations aim to explain the results
of the detailed analysis of the experimental data for $^{230}$Th.
%%%%%%%%%%%%%%%%%%%%%%%%%%%%%%%%%%%%%%%%%%%%%%%%%%%%%%%%%%%%%%%%%%%%%%%
\begin{figure}[!]
\begin{center}
\epsfig{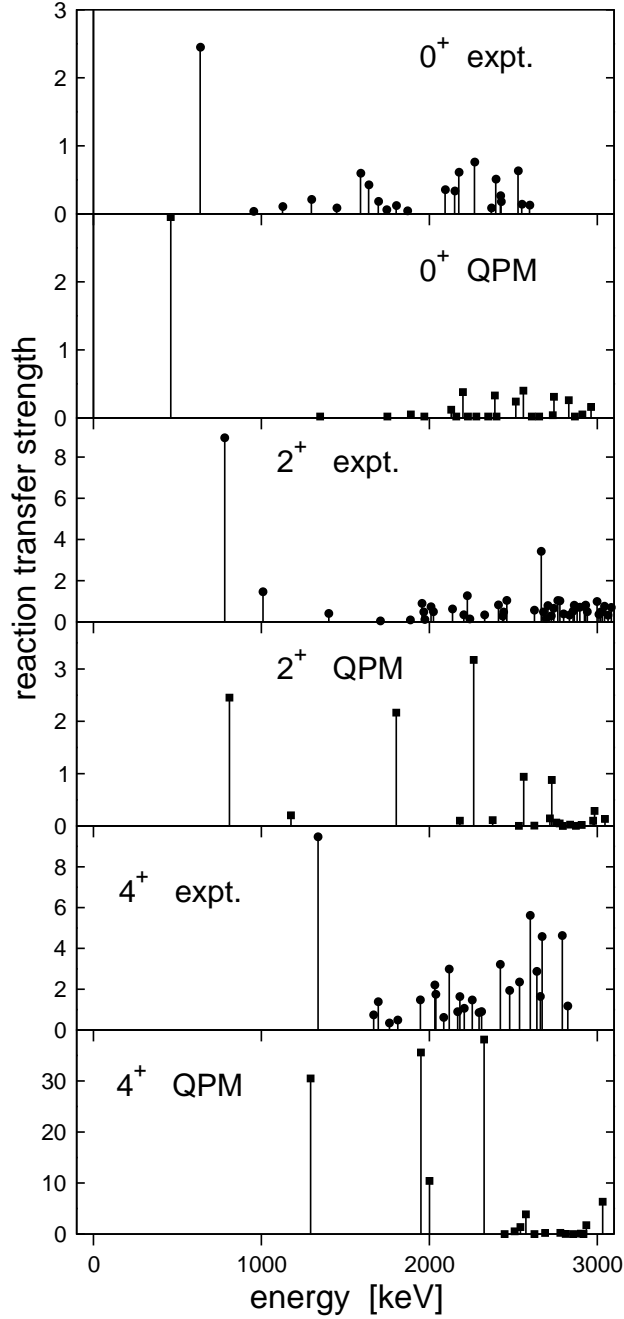}
    \caption{\label{fig:compil-zeroQPM} Comparison of experimental and calculated
    (QPM) 0$^+$, 2$^+$ and 4$^+$  relative level  reaction  strengths for the (p,t)
    reaction. The 2$^+$ and 4$^+$ states assumed to belong to rotational bands
    (see Sec.~\ref{sec:bands}) are not included. The experimental strengths for
    0$^+_{g.s.}$, 2$^+_1$ and 4$^+_1$  being rotational states are normalized to 10
    (the latter two are not shown in figure).}
\end{center}
\end{figure}
%%%%%%%%%%%%%%%%%%%%%%%%%%%%%%%%%%%%%%%%%%%%%%%%%%%%%%%%%%%%%%%%%%%%%%

In the QPM \cite{Sol92}  the Hamiltonian in a separable

generalized form is adopted to generate  the quasiparticle RPA
phonons described by the operators
\begin{equation}\label{eq:phonon}
Q^{\dag}_{i \nu}=\frac{1}{2} \sum_{q_1 q_2}
 ( \psi^{i \nu}_{q_1 q_2}
\alpha^{\dag}_{q_1} \alpha^{\dag}_{q_2} - \phi^{i\nu}_{q_1
 q_2}\alpha_{q_2} \alpha_{q_1})
\end{equation}
The Hamiltonian expressed in terms of these phonon operators is
diagonalized in the space spanned by one- and two-phonon states.
The QPM eigenstates have the structure
\begin{equation}\label{eq:psi-func}
\Psi_{nK} = \sum_iC_i^{(n)}Q^{\dag}_{i \lambda K}|0 \rangle +
\sum_{v_1v_2}C^{(n)}_{v_1v_2}[Q^{\dag}_{v_1} \otimes
Q^{\dag}_{v_2}]_K |0 \rangle,
\end{equation}
where $\lambda K$ label the multipolarity and magnetic component
of the phonon operator. Each of these states represents the
intrinsic component of the total wave function
\begin{eqnarray}\label{eq:total}
\Psi_{nMK}^I = \sqrt{\frac{(1+\delta_{K0})(2I+1)}{16\pi^2}}\\
\nonumber \times [D^I_{MK}\Psi_{nK} +
(-)^{I+K}D^I_{M-K}\Psi_{nK}],
\end{eqnarray}
where $D^I_{MK}$ is the Wigner matrix. No free parameters are used
in these calculations, all  physical input quantities and
constants are determined by an independent fit to the experimental
data in neighboring odd nuclei. For details see \cite{Lo05}.
%%%%%%%%%%%%%%%%%%%%%%%%%%%%%%%%%%%%%%%%%%%%%%%%%%%%%%%%%%%%%%%%%%%%%%
\begin{figure}[!]
\begin{center}
\epsfig{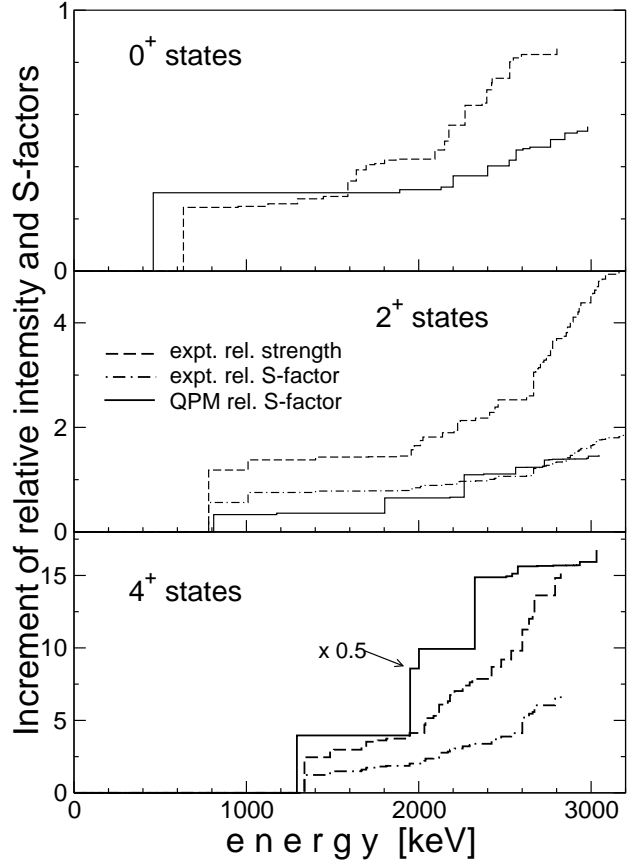}
    \caption{\label{fig:increment} Experimental increments of the (p,t)
    strength in comparison with the QPM calculations. The 2$^+$ and 4$^+$
    states assumed as belonging to rotational bands
    (see Sec.~\ref{sec:bands}) are not included.}
\end{center}
\end{figure}
%%%%%%%%%%%%%%%%%%%%%%%%%%%%%%%%%%%%%%%%%%%%%%%%%%%%%%%%%%%%%%%%%%%%%

The experimental spectra of the $0^{+}$, 2$^+$ and 4$^+$ states in
$^{230}$Th are compared with the results of the QPM calculations
in Fig.~\ref{fig:compil-zeroQPM}.  The QPM considers  only
vibrational 2$^+$ and 4$^+$ excitations  in the even-even nuclei,
therefore all other excitations have to be excluded from the
comparison.  Only the 2$^+$ and 4$^+$ states not belonging to
rotational bands (see Sec.~\ref{sec:bands}) are included in
Fig.~\ref{fig:compil-zeroQPM}, i.e. these states are assumed to be
mainly of vibrational structure. The QPM generates 23 $0^+$ states
below 2.8 MeV in fair agreement with the 24 identified states. At
the same time the QPM yields less 2$^+$ and 4$^+$ excited states
than the ones observed experimentally. The numbers of 2$^+$ and
4$^+$ states generated up to 3 MeV are 20 and 17, respectively,
considerably less than the observed 50 and 27 corresponding
experimental levels (Fig.~\ref{fig:compil-zeroQPM}). It seems that
taking into account the two-quasiparticle and  pairing vibrational
states not excluded from consideration (there are problems in
their identification) cannot explain this large difference. Beside
the ground state, experiment reveals one very strong peak for
every of 0$^+$, 2$^+$ and 4$^+$ state and small strengths for
other states. The calculation correctly yields strong peaks close
in magnitude and position to the experimental ones for the 0$^+$,
2$^+$ and 4$^+$ states. These strong peaks form the suggested
multiplet shown in Fig.~\ref{fig:quadruplet}. The calculation
correctly yields small strengths for other peaks in the case of
0$^+$ excitations. At the same time, besides the first strong
peaks, the QPM predicts for the 2$^+$ and 4$^+$ states two other
strong peaks not observed experimentally.

In contrast to the {\it spdf}-IBM, the QPM  is able to reproduce
the two-neutron transfer strength. The wave functions
(\ref{eq:total}) can be used to compute the (p,t) normalized
transfer spectroscopic factors
\begin{equation}\label{spec-fact}
S_n(p,t) = \left[\frac{\Gamma_n(p,t)}{\Gamma_0(p,t)}\right]^2,
\end{equation}
where the amplitudes are given by \cite{Bro69}
\begin{equation}\label{spec-amplit}
\Gamma_n(p,t) = \langle \Psi^I_{nMK},N-2|\sum_{q_1q_2} r^I
Y_{IK}a_{q_1}a_{q_2}|\Psi_0,N\rangle.
\end{equation}
The amplitude $\Gamma_0(p,t)$ refers to the transitions to the $I$
members of the ground state rotational band \cite{Lo05}.
 In Fig.~\ref{fig:increment} we present the increments
 of the (p,t) strength  to the 0$^+$, 2$^+$ and 4$^+$ states
 and that of the spectroscopic factors  derived from the DWBA analysis
 for these states. They are given relative to those for corresponding
 states of the ground state band and are compared with the calculated
 normalized spectroscopic factors.
Since the DWBA analysis for the 0$^+$ states included
 different configurations of the transferred neutrons, only the (p,t) strength
 ratio is given in Fig.~\ref{fig:increment} for these states. For
 other states only the $(2g_{9/2})^2$ neutron configuration was
 accepted. According to the DWBA formalism in the case of the same
 neutron configuration and the direct one-step transfer the strength
 ratios have to be close to the spectroscopic factor ratios.
 This is not the case for $^{230}$Th, since a considerable difference of
 these two ratios is observed.

 As one can see, the calculations of $0^+$ for the (p,t) strength ratio
 are in fair agreement with the experiment. For the $2^+$ states the
 calculations are in good agreement with the measured spectroscopic
 factor ratios, however
 almost two times smaller than the measured strength ratios.
 We have to note that the (p,t) strengths for the 0$^+$ and 2$^+$
 states not firmly assigned are small and do not influence
 considerably the results of comparison . At the same time withdrawal
 of these states from consideration will only improve the agreement
 with the calculations.
 For the 4$^+$ states the calculated spectroscopic
 factors are more than two times larger than the experimental strength
 and spectroscopic factor ratios. The difference possibly stems from
 the change of the (p,t) cross section caused by the inclusion of
 additional two-step paths of the neutron transfer for the $2^+$
 and 4$^+$ members of the ground state band. As one can see from
 Fig.~\ref{fig:angl_distr_natur}, the angular distributions for
 these states differ considerably from those for direct one-step
 transfer and can be fitted only by an inclusion of a two-step excitation
 number which is the largest for the 4$^+$ state. For the 0$^+$
 states, where only the one-step transfer is possible, the agreement
 between calculation and experiment is good. For both 2$^+$ and
 4$^+$ states the energies at which the experimental strength first
 appears agrees with the calculated ones. This means that the energies of the
 first vibrational 2$^+$ and 4$^+$ excitations predicted by the
 QPM are in fair agreement with the experiment (recall that the states
 belonging to the rotational bands are excluded from this
 comparison).

The nature of $0^+$ excitations as well as for $2^+$ and 4$^+$
states in the QPM differs greatly from that in the {\it spdf}-IBM.
In all low-lying states  quadrupole phonons are dominant and the
octupole phonons are predicted to play a relatively modest role.
This might be acceptable in the vibration-like $^{230}$Th,  but
the same is predicted for the octupole soft $^{228}$Th. The
spectrum is explained from the QPM calculation procedure (Pauli
principle) as a redistribution of the strength of the lowest
two-octupole phonons among many closely packed QPM $0^+$ states
\cite{Lo05}. To assess this aspect, calculations for octupole
deformed lighter isotopes of Th would be important.

Other predictions of the QPM can be tested experimentally only for
the lowest states. The measured  values of $B(E2;0^+_1 \rightarrow
2^+_0) = 1.1$ W.u. is close to the computed value of 1.7 W.u.  At
the same time the calculated monopole transition strength
$\rho^2(E0;0^+_1 \rightarrow 0^+_g)=1.48 \cdot 10^{-3}$  is two
orders of magnitude smaller than the  experimental value of
$126(13) \cdot 10^{-3}$ \cite{Ack94}. Again, we would like to
stress the necessity of systematic measurements of electromagnetic
properties of higher excited $0^+$ states to understand their
nature.

%%%%%%%%%%%%%%%%%%%%%%%%%%%%%%%%%%%%%%%%%%%%%%%%%%%%%%%%%%%%%%%%%%%%

\section{conclusion}

Excited states in $^{230}$Th have been studied in (p,t) transfer
reactions. About 200 levels were assigned using a DWBA fit
procedure. Among them, 24 excited $0^+$ states have been found in
this nucleus, most of them have not been experimentally observed
previously. Their accumulated strength makes up more than 70\,\%
of the ground state strength. Firm assignments have been made for
most of the 2$^+$ and 4$^+$  states and for some of the 6$^+$
states. These assignments allowed the identification of sequences
of states which have the features of rotational bands with
definite inertial parameters. The 2$^+$, 4$^+$ and 6$^+$ states
not included in these bands have been considered as vibration-like
excitations. Multiplets of states are suggested which can be
treated as one- and two-phonon octupole quadruplets. The
experimental data are compared with  {\it spdf}-IBM and QPM
calculations. Giving an approximately correct number of $0^+$
states, these models provide different predictions for the
structure of these states. They are also in conflict with the
apparent structure of the states inferred from the moments of
inertia of the rotational bands built on them. More specifically,
as follows from the moments of inertia, the 0$^+$ states have a
different intrinsic structures, which is in contradiction with the
predictions of both the IBM (predominantly octupole bands) and the
QPM (predominantly quadrupole bands). A remarkable feature of the
QPM is the prediction of strong first vibrational excitations
close in magnitude and position to the experimental ones. The
numbers of 2$^+$ and 4$^+$ states are underestimated by both
theories. Spectroscopic factors from the (p,t) reaction, and the
trend in their change with the excitation energy, are
approximately reproduced by the QPM for the 0$^+$ and 2$^+$ states
and overestimated by  theory for the 4$^+$ states. The lack of
additional information does not allow  for final conclusions on
the validity of the theoretical approaches. Therefore we hope that
our new data will stimulate further experimental and theoretical
studies. Accurate experiments and a detailed analysis similar to
the present work are desirable for other nuclei in this region.
Challenging experiments on gamma spectroscopy following (p,t)
reactions would give much needed information.

%%%%%%%%%%

\section*{\boldmath ACKNOWLEDGEMENT}

\noindent We thank H.J.~Maier for the preparation of the targets.
The work was supported by the DFG (C4-Gr894/2-3, Gu179/3,
Jo391/2-3), MLL, and  US-DOE, contract number DE-FG02-91ER-40609.

%%%%%%%%%%%%%%%%%%%%%%%%%%%%%%%%%%%%%%%%%%%%%%%%

\vspace{20mm}

{\Large \bf APPENDIX: Angular distributions and their fit by the
CHUCK3
         calculations for all excited states observed in the
         present study}\\

%%%%%%%%%%%%%%%%%%%%%%%%%%%%%%%%%%%%%%%%%%%%%%%%%%%%%%%%%%%%%%%%%%%%%%%
\psfig{file=fig15a.eps, width=16cm,angle=0}
%%%%%%%%%%%%%%%%%%%%%%%%%%%%%%%%%%%%%%%%%%%%%%%%%%%%%%%%%%%%%%%%%%%%%%

%%%%%%%%%%%%%%%%%%%%%%%%%%%%%%%%%%%%%%%%%%%%%%%%%%%%%%%%%%%%%%%%%%%%%%
\psfig{file=fig15b.eps, width=16cm,angle=0}
%%%%%%%%%%%%%%%%%%%%%%%%%%%%%%%%%%%%%%%%%%%%%%%%%%%%%%%%%%%%%%%%%%%%%%

%%%%%%%%%%%%%%%%%%%%%%%%%%%%%%%%%%%%%%%%%%%%%%%%%%%%%%%%%%%%%%%%%%%%%%
\psfig{file=fig15c.eps, width=16cm,angle=0}
%%%%%%%%%%%%%%%%%%%%%%%%%%%%%%%%%%%%%%%%%%%%%%%%%%%%%%%%%%%%%%%%%%%%%%

%%%%%%%%%%%%%%%%%%%%%%%%%%%%%%%%%%%%%%%%%%%%%%%%%%%%%%%%%%%%%%%%%%%%%%
\psfig{file=fig15d.eps, width=16cm,angle=0}
%%%%%%%%%%%%%%%%%%%%%%%%%%%%%%%%%%%%%%%%%%%%%%%%%%%%%%%%%%%%%%%%%%%%%%

%%%%%%%%%%%%%%%%%%%%%%%%%%%%%%%%%%%%%%%%%%%%%%%%%%%%%%%%%%%%%%%%%%%%%%
\psfig{file=fig15e.eps, width=16cm,angle=0}
%%%%%%%%%%%%%%%%%%%%%%%%%%%%%%%%%%%%%%%%%%%%%%%%%%%%%%%%%%%%%%%%%%%%%%

%%%%%%%%%%%%%%%%%%%%%%%%%%%%%%%%%%%%%%%%%%%%%%%%%%%%%%%%%%%%%%%%%%%%%%
\psfig{file=fig15f.eps, width=16cm,angle=0}
%%%%%%%%%%%%%%%%%%%%%%%%%%%%%%%%%%%%%%%%%%%%%%%%%%%%%%%%%%%%%%%%%%%%%%

%%%%%%%%%%%%%%%%%%%%%%%%%%%%%%%%%%%%%%%%%%%%%%%%%%%%%%%%%%%%%%%%%%%%%%
\psfig{file=fig15g.eps, width=16cm,angle=0}
%%%%%%%%%%%%%%%%%%%%%%%%%%%%%%%%%%%%%%%%%%%%%%%%%%%%%%%%%%%%%%%%%%%%%%

%%%%%%%%%%%%%%%%%%%%%%%%%%%%%%%%%%%%%%%%%%%%%%%%%%%%%%%%%%%%%%%%%%%%%%
\psfig{file=fig15h.eps, width=16cm,angle=0}
%%%%%%%%%%%%%%%%%%%%%%%%%%%%%%%%%%%%%%%%%%%%%%%%%%%%%%%%%%%%%%%%%%%%%%

\end{document}